\documentclass[12pt]{article}
\usepackage{epsfig}
\usepackage{amsmath}
\usepackage{hhline}
\usepackage{amssymb}
\usepackage{times}
\usepackage{cite}
\usepackage{bold-extra}

\newlength{\dinwidth}
\newlength{\dinmargin}
\begin{document}  
\newcommand{\pom}{{I\!\!P}}
\newcommand{\reg}{{I\!\!R}}
\newcommand{\slowpi}{\pi_{\mathit{slow}}}
\newcommand{\fiidiii}{F_2^{D(3)}}
\newcommand{\fiidiiiarg}{\fiidiii\,(\beta,\,Q^2,\,x)}
\newcommand{\n}{1.19\pm 0.06 (stat.) \pm0.07 (syst.)}
\newcommand{\nz}{1.30\pm 0.08 (stat.)^{+0.08}_{-0.14} (syst.)}
\newcommand{\fiidiiiful}{F_2^{D(4)}\,(\beta,\,Q^2,\,x,\,t)}
\newcommand{\fiipom}{\tilde F_2^D}
\newcommand{\ALPHA}{1.10\pm0.03 (stat.) \pm0.04 (syst.)}
\newcommand{\ALPHAZ}{1.15\pm0.04 (stat.)^{+0.04}_{-0.07} (syst.)}
\newcommand{\fiipomarg}{\fiipom\,(\beta,\,Q^2)}
\newcommand{\pomflux}{f_{\pom / p}}
\newcommand{\nxpom}{1.19\pm 0.06 (stat.) \pm0.07 (syst.)}
\newcommand {\gapprox}
   {\raisebox{-0.7ex}{$\stackrel {\textstyle>}{\sim}$}}
\newcommand {\lapprox}
   {\raisebox{-0.7ex}{$\stackrel {\textstyle<}{\sim}$}}
\def\gsim{\,\lower.25ex\hbox{$\scriptstyle\sim$}\kern-1.30ex%
\raise 0.55ex\hbox{$\scriptstyle >$}\,}
\def\lsim{\,\lower.25ex\hbox{$\scriptstyle\sim$}\kern-1.30ex%
\raise 0.55ex\hbox{$\scriptstyle <$}\,}
\newcommand{\pomfluxarg}{f_{\pom / p}\,(x_\pom)}
\newcommand{\dsf}{\mbox{$F_2^{D(3)}$}}
\newcommand{\dsfva}{\mbox{$F_2^{D(3)}(\beta,Q^2,x_{I\!\!P})$}}
\newcommand{\dsfvb}{\mbox{$F_2^{D(3)}(\beta,Q^2,x)$}}
\newcommand{\dsfpom}{$F_2^{I\!\!P}$}
\newcommand{\gap}{\stackrel{>}{\sim}}
\newcommand{\lap}{\stackrel{<}{\sim}}
\newcommand{\fem}{$F_2^{em}$}
\newcommand{\tsnmp}{$\tilde{\sigma}_{NC}(e^{\mp})$}
\newcommand{\tsnm}{$\tilde{\sigma}_{NC}(e^-)$}
\newcommand{\tsnp}{$\tilde{\sigma}_{NC}(e^+)$}
\newcommand{\st}{$\star$}
\newcommand{\sst}{$\star \star$}
\newcommand{\ssst}{$\star \star \star$}
\newcommand{\sssst}{$\star \star \star \star$}
\newcommand{\tw}{\theta_W}
\newcommand{\sw}{\sin{\theta_W}}
\newcommand{\cw}{\cos{\theta_W}}
\newcommand{\sww}{\sin^2{\theta_W}}
\newcommand{\cww}{\cos^2{\theta_W}}
\newcommand{\trm}{m_{\perp}}
\newcommand{\trp}{p_{\perp}}
\newcommand{\trmm}{m_{\perp}^2}
\newcommand{\trpp}{p_{\perp}^2}
\newcommand{\alp}{\alpha_s}

\newcommand{\alps}{\alpha_s}
\newcommand{\sqrts}{$\sqrt{s}$}
\newcommand{\LO}{$O(\alpha_s^0)$}
\newcommand{\Oa}{$O(\alpha_s)$}
\newcommand{\Oaa}{$O(\alpha_s^2)$}
\newcommand{\PT}{p_{\perp}}
\newcommand{\JPSI}{J/\psi}
\newcommand{\sh}{\hat{s}}
\newcommand{\uh}{\hat{u}}
\newcommand{\MP}{m_{J/\psi}}
\newcommand{\PO}{I\!\!P}
\newcommand{\xbj}{x}
\newcommand{\xpom}{x_{\PO}}
\newcommand{\ttbs}{\char'134}
\newcommand{\xpomlo}{3\times10^{-4}}  
\newcommand{\xpomup}{0.05}  
\newcommand{\dgr}{^\circ}
\newcommand{\pbarnt}{\,\mbox{{\rm pb$^{-1}$}}}
\newcommand{\gev}{\,\mbox{GeV}}
\newcommand{\WBoson}{\mbox{$W$}}
\newcommand{\fbarn}{\,\mbox{{\rm fb}}}
\newcommand{\fbarnt}{\,\mbox{{\rm fb$^{-1}$}}}
\newcommand{\dsdx}[1]{${\rm d}\sigma/{\rm d} #1\,$}
\newcommand{\eV}{\mbox{e\hspace{-0.08em}V}}
%
%
\newcommand{\qsq}{\ensuremath{Q^2} }
\newcommand{\gevsq}{\ensuremath{\mathrm{GeV}^2} }
\newcommand{\et}{\ensuremath{E_t^*} }
\newcommand{\rap}{\ensuremath{\eta^*} }
\newcommand{\gp}{\ensuremath{\gamma^*}p }
\newcommand{\dsiget}{\ensuremath{{\rm d}\sigma_{ep}/{\rm d}E_t^*} }
\newcommand{\dsigrap}{\ensuremath{{\rm d}\sigma_{ep}/{\rm d}\eta^*} }

\newcommand{\dstar}{\ensuremath{D^*}}
\newcommand{\dstarp}{\ensuremath{D^{*+}}}
\newcommand{\dstarm}{\ensuremath{D^{*-}}}
\newcommand{\dstarpm}{\ensuremath{D^{*\pm}}}
\newcommand{\zDs}{\ensuremath{z(\dstar )}}
\newcommand{\Wgp}{\ensuremath{W_{\gamma p}}}
\newcommand{\ptds}{\ensuremath{p_T(\dstar )}}
\newcommand{\etads}{\ensuremath{\eta(\dstar )}}
\newcommand{\ptj}{\ensuremath{p_T(\mbox{jet})}}
\newcommand{\ptjn}[1]{\ensuremath{p_T(\mbox{jet$_{#1}$})}}
\newcommand{\etaj}{\ensuremath{\eta(\mbox{jet})}}
\newcommand{\detadsj}{\ensuremath{\eta(\dstar )\, \mbox{-}\, \etaj}}
\newcommand{\rnorm}{\ensuremath{R^{\rm norm}}}
\newcommand{\shat}{\ensuremath{\hat{s}}}
\newcommand{\dstarpj}{\ensuremath{\dstar\! \mbox{+jet}}}
\newcommand{\dstarjet}{\ensuremath{\dstar\ \mbox{jet}}}
\newcommand{\dstardj}{{\it \dstar\ tagged dijet}}
\newcommand{\dstarDj}{\dstar-tagged dijet}
\newcommand{\otherj}{{\ensuremath{\mbox{other }\mbox{jet}}}}
\newcommand{\dstarpotherj}{{\it \ensuremath{\dstar\! \mbox{ + other }\mbox{jet}}}}
\newcommand{\dstarpotherJ}{\ensuremath{\dstar\! \mbox{ + other }\mbox{Jet}}}
\newcommand{\pt}{\ensuremath{p_T}}
\newcommand{\kt}{\ensuremath{k_t}}
\newcommand{\dphidsj}{\ensuremath{|\Delta \varphi|}}
\newcommand{\xgjj}{\ensuremath{x_\gamma}}
\newcommand{\as}{\ensuremath{\alpha_s}}
\def\pythia{{\scshape Pythia}}
\def\cascade{{\scshape Cascade}}
\def\PYTHIA{{\scshape Pythia}}
\def\CASCADE{{\scshape Cascade}}

\newcommand{\grad}{\ensuremath{^\circ\!}} 

\def\Journal#1#2#3#4{{#1} {\bf #2} (#3) #4}
\def\NCA{Nuovo Cimento}
\def\NIM{Nucl. Instrum. Methods}
\def\NIMA{{Nucl. Instrum. Methods} {\bf A}}
\def\NPB{{Nucl. Phys.}   {\bf B}}
\def\PLB{{Phys. Lett.}   {\bf B}}
\def\PRL{ Phys. Rev. Lett.}
\def\PRD{{Phys. Rev.}	 {\bf D}}
\def\ZPC{{Z. Phys.}	 {\bf C}}
\def\EPJC{{Eur. Phys. J.} {\bf C}}
\def\CPC{Comp. Phys. Commun.}

\begin{titlepage}

\noindent
\begin{flushleft}
{\tt DESY 11-248    \hfill    ISSN 0418-9833} \\
{\tt December 2011}                  \\
\end{flushleft}

\noindent

\vspace{2cm}
\begin{center}
\begin{Large}

{\bf \boldmath Measurement of Inclusive and Dijet \dstar\ Meson 
Cross Sections in Photoproduction at HERA}

\vspace{2cm}

H1 Collaboration

\end{Large}
\end{center}

\vspace{2cm}

\begin{abstract}
The inclusive photoproduction of \dstar\ mesons
and of {\dstarDj}s is investigated with the H1 detector at 
the $ep$ collider HERA. 
The kinematic region covers 
small photon virtualities $Q^{2} < 2\ {\rm GeV}^2$ and photon-proton 
centre-of-mass energies of $100 < W_{\gamma p} < 285\ {\rm GeV}$. 
Inclusive \dstar\ meson differential cross sections are measured for central 
rapidities $|\eta(\dstar)| < 1.5$ and transverse momenta
$p_{T}(\dstar) >1.8\ {\rm GeV}$. 
The heavy quark production process is further investigated in
events with at least two jets with transverse momentum
$\ptj > 3.5\ {\rm GeV}$ each, one containing the \dstar\ meson.
Differential cross sections for \dstarDj\ production and for
correlations between the jets are measured in the range
$|\eta(\dstar)| < 1.5$ and $p_{T}(\dstar) >2.1\ {\rm GeV}$.
The results are compared with predictions from Monte Carlo simulations
and next-to-leading order perturbative QCD calculations.
\end{abstract}

\vspace{1.5cm}

\begin{center}
Submitted to Eur. Phys. J. {\bf C}
\end{center}

\end{titlepage}

\begin{flushleft}

F.D.~Aaron$^{5,48}$,           
C.~Alexa$^{5}$,                
V.~Andreev$^{25}$,             
S.~Backovic$^{30}$,            
A.~Baghdasaryan$^{38}$,        
S.~Baghdasaryan$^{38}$,        
E.~Barrelet$^{29}$,            
W.~Bartel$^{11}$,              
K.~Begzsuren$^{35}$,           
A.~Belousov$^{25}$
% ...

\bigskip{\it
% ...
$ ^{5}$ National Institute for Physics and Nuclear Engineering (NIPNE) ,
          Bucharest, Romania$^{ m}$ \\
% ...
$ ^{11}$ DESY, Hamburg, Germany \\
% ...
$ ^{25}$ Lebedev Physical Institute, Moscow, Russia$^{ e}$ \\
% ...
$ ^{29}$ LPNHE, Universit\'e Pierre et Marie Curie Paris 6,
          Universit\'e Denis Diderot Paris 7, CNRS/IN2P3, Paris, France \\
 $ ^{30}$ Faculty of Science, University of Montenegro,
          Podgorica, Montenegro$^{ n}$ \\
% ...
$ ^{35}$ Institute of Physics and Technology of the Mongolian
          Academy of Sciences, Ulaanbaatar, Mongolia \\
% ...
$ ^{38}$ Yerevan Physics Institute, Yerevan, Armenia \\
% ...
$ ^{48}$ Also at Faculty of Physics, University of Bucharest,
          Bucharest, Romania \\
% ...
$ ^e$ Supported by the Deutsche Forschungsgemeinschaft \\
% ...
$ ^m$ Supported by the Romanian National Authority for Scientific Research
      under the contract PN 09370101 \\
 $ ^n$ Partially Supported by Ministry of Science of Montenegro,
      no. 05-1/3-3352 \\
}

\end{flushleft}

\newpage
\section{Introduction}
At the electron-proton collider HERA charm quarks are predominantly 
produced via boson gluon fusion, $\gamma g \rightarrow c \bar{c}$,
where the photon is emitted from the incoming lepton and the
gluon originates from the proton. The cross section is largest 
for photoproduction, i.e.~for photons with negative four-momentum squared 
(virtuality) $Q^2 \simeq 0\ {\rm GeV}^2$. In addition to 
hard direct scattering off the photon,
processes have to be considered in which 
the partonic structure of the photon is resolved.
The charm quark mass provides a hard scale which 
justifies the applicability of perturbative QCD (pQCD).
 
Previous measurements of the photoproduction of charm quarks at HERA 
cover inclusive \dstar\ meson 
production~\cite{zeusdstar98,h1dstar98,h1dstar06}, production 
of \dstar\ mesons with associated dijets~\cite{zeusdstar98,zeusdstarjets03,
zeusdstarjets05,h1dstar06} and heavy quark production 
using events with a \dstar\ meson and a muon~\cite{h1dstarmu}. 
In this paper, single and double differential cross sections are presented
for the inclusive production of \dstar\ mesons and the production
of two jets with one of the jets containing the \dstar\ meson. 
They are compared to leading and next-to-leading order pQCD predictions
using different hadronisation models.
Compared to the previous H1 analysis of inclusive \dstar\ 
photoproduction~\cite{h1dstar06}, a seven times larger signal sample 
is analysed here. 

Studying events in which at least two jets could be reconstructed, 
with one of the jets
containing the \dstar\ meson, allows further investigations of the
details of the heavy quark production process.
The jets are measured down to transverse momenta of $\ptj = 3.5\ {\rm GeV}$.
While the jet containing the \dstar\ meson originates from a charm or 
anticharm quark produced in the hard subprocess, the
non-\dstar -tagged jet, refered to as {\it \otherj}, can result from 
either the other heavy quark or a light parton (e.g.~a gluon). 
Correlations between the two jets are studied using variables which are 
sensitive to higher order effects and to the
longitudinal as well as to the transverse momentum components of the partons
entering the hard scattering process.

\section{QCD Calculations}

The data presented in this analysis are compared with Monte Carlo 
simulations based on leading order (LO) matrix elements
supplemented by parton showers and with next-to-leading 
order (NLO) calculations. The calculations are performed using either the 
collinear factorisation or the $k_t$-factorisation approach.
The collinear factorisation makes use of the DGLAP~\cite{DGLAP} evolution 
equations, while in $k_t$-factorisation the CCFM~\cite{CCFM} evolution
equations are employed. In the collinear approach transverse momenta obtained 
through the initial 
state QCD evolution are neglected and the transverse momenta are 
generated in the hard scattering process. Effects from the non-vanishing
transverse momenta of the gluons enter only at the NLO level. In the 
$k_t$-factorisation ansatz the transverse momenta of incoming 
gluons, $k_t$, are already included at leading order both in the 
off-shell matrix element and the $k_t$-dependent unintegrated gluon 
density~\cite{uPDF}. Corrections appearing only at higher order 
in collinear factorisation are hence partially included at LO in the 
$k_t$-factorisation approach.

For charm quark photoproduction two classes of processes occur, 
the direct-photon and the resolved-photon processes. In the direct 
processes the photon emitted from the beam lepton enters directly the hard 
interaction, 
whereas in the resolved processes the photon acts as the source of incoming 
partons, one of which takes part in the hard interaction. The distinction 
between these two classes depends on the factorisation scheme and the 
order in which the calculation is performed.

The production of heavy quarks is calculated either in the massive 
scheme, where heavy quarks are produced only perturbatively 
via boson gluon fusion, or in the massless scheme, where
heavy quarks are treated as massless partons. These two schemes are expected to 
be appropriate in different regions of phase space~\cite{Tung}:
the massive scheme is expected to be reliable when the transverse momentum 
$p_{T}$ of the heavy quark is of similar size compared to the charm 
mass $m_{c}$, whereas the massless scheme is expected to be valid
for $p_{T} \gg m_{c}$. 
In the general-mass variable-flavour-number scheme (GMVFNS) a smooth 
transition from the massive to the massless scheme is 
provided.  
The structure of the proton and of the photon are described by
parton distribution functions (PDFs), that have been determined by
fits to data in various heavy flavour schemes and at different orders
of pQCD. 

Monte Carlo (MC) generators are used to simulate detector
effects in order to determine the acceptance and the efficiency for 
selecting events and to estimate the systematic uncertainties
associated with the measurement. The generated events are passed 
through a detailed simulation of the detector response based on the GEANT 
simulation programm~\cite{geant} and are processed
using the same reconstruction and analysis chain as is used for the data.
The following two MC generators are used:

\begin{description}
\item[\PYTHIA:]
The MC program \PYTHIA~\cite{pythia} is based on LO QCD matrix elements
with leading-log parton showers in the collinear factorisation approach.
\PYTHIA\ includes both direct photon gluon fusion and resolved-photon 
processes. In the resolved-photon processes either a charm quark or a 
gluon from the photon enters the hard scattering. 
In the inclusive mode of \PYTHIA\ used here charm quarks are treated 
as massless partons in all steps of the calculation in both types of processes.
The hadronisation process is simulated using the Lund string fragmentation 
model~\cite{lundstring}. The Bowler fragmentation model~\cite{bowler} is 
applied to fragment the charm quark into a \dstar\ meson. 
The longitudinal part of the fragmentation is reweighted 
to the parameterisation by Kartvelishvili et al.~\cite{kartvelish} which
depends on a single parameter $\alpha$. 
The latter is set to the values 
determined by H1~\cite{Aaron:2008tt}, which depend on the centre-of-mass 
energy squared of the hard subprocess $\hat{s}$ (see table~\ref{fragpar}). 
The proton structure is described by the
PDF set CTEQ6L~\cite{cteq6l}. 
For the photon the PDF set GRV-G LO~\cite{grvlo} is used. 

\renewcommand{\arraystretch}{1.15} 
\begin{table}[ht]
\begin{center}
\begin{tabular}{|l||c||c|c||c|c|}
\hline
\multicolumn{6}{|c|}{ }\\[-0.4cm]
\multicolumn{6}{|c|}{\bf Fragmentation parameter \boldmath$\alpha$\unboldmath}\\[0.1cm]\hline
 & & \multicolumn{2}{c||}{\bf \PYTHIA} & \multicolumn{2}{c|}{\bf \CASCADE}\\\hline
 & $\shat_{threshold}$ & 
$\alpha$ for & $\alpha$ for  & 
$\alpha$ for & $\alpha$ for  \\ 
 & $[{\rm GeV}^2]$ & 
$\shat < \shat_{threshold}$ & $\shat \ge \shat_{threshold}$ & 
$\shat < \shat_{threshold}$ & $\shat \ge \shat_{threshold}$ \\ \hline\hline
Central value & $70$ & $10.3$ & $4.4$ & $8.4$ & $4.5$ \\ \hline
Variations & $70$ & $\phantom{1}8.7$ & $3.9$ & $7.3$ & $3.9$ \\
 & $70$ & $12.2$ & $5.0$ & $9.8$ & $5.1$ \\
 & $50$ & $10.3$ & $4.4$ & $8.4$ & $4.5$ \\ 
 & $90$ & $10.3$ & $4.4$ & $8.4$ & $4.5$ \\ \hline
\end{tabular}

\caption{\label{fragpar}
Fragmentation parameters $\alpha$ in the Kartvelishvili 
parameterisation used in the MC simulations.
In the two regions of the invariant mass squared of the
$c\bar{c}$ pair, $\shat$, separated by the boundary $\shat_{threshold}$,
two different values of $\alpha$ are used.
}
\end{center}
\end{table}
\renewcommand{\arraystretch}{1.0}

\item[\CASCADE:] 
The \CASCADE~\cite{cascade} MC program is used for simulating events 
based on LO QCD calculations in the
$k_t$-factorisation approach. The direct boson gluon fusion process is 
implemented using off-shell matrix elements and incoming gluons which can 
have non-vanishing transverse momenta. Higher order QCD corrections are 
simulated with initial state parton showers applying the CCFM 
evolution~\cite{CCFM}. The unintegrated PDFs of the proton from 
set A0~\cite{a0} are used. The 
hadronisation of partons is performed with the Lund string model as 
implemented in \PYTHIA. For the fragmentation of the charm quarks into 
\dstar\ mesons the same reweighting procedure to the  
parameterisation of Kartvelishvili et al.~is applied as in the case
of \PYTHIA.

\end{description}

For the comparison of data with NLO predictions, calculations 
based on the massive approach and the general mass variable flavor number 
scheme are used.
The uncertainties of the calculations are estimated by varying the charm 
mass, $m_c$, the factorisation scale, $\mu_f$, and the renormalisation 
scale, $\mu_r$. The detailed 
settings are given in table~\ref{Tab_NLO_parameters}.
For the comparison in the \dstarDj\ sample only MC@NLO is used since it 
provides a full hadronisation of the final state.

 
\renewcommand{\arraystretch}{1.15} 
\begin{table}[ht]
\begin{center}
\begin{tabular}{|l||c|c c||c|c c||c|c c|}
\hline
 & \multicolumn{3}{c||}{\bf FMNR} & \multicolumn{3}{c||}{\bf GMVFNS} & 
\multicolumn{3}{c|}{\bf MC@NLO} \\[0.1cm]\hline
Parameter  & Central & \multicolumn{2}{c||}{Variations} 
& Central & \multicolumn{2}{c||}{Variations} 
& Central & \multicolumn{2}{c|}{Variations} \\\hline\hline
Charm mass $m_c/{\rm GeV}$ & $1.5$ & $1.3$ & $1.7$ 
& $1.5$ & & & $1.5$ & $1.3$ & $1.7$ \\
Renorm. Scale $\mu_{r}/m_T$ & $1$ &  $0.5$ & $2$
& $1$ & $0.5$ & $2$ & $1$ & $0.5$ & $2$ \\
Fact. Scale $\mu_{f}/m_T$ & $2$ &  $1$ & $4$
& $1$ & $0.5$ & $2$ & $2$ & $1$ & $4$ \\
\hline
\end{tabular}
\caption{
Parameters and variations used in the NLO calculations of FMNR~\cite{Frixione,FMNR},
GMVFNS~\cite{Kramer, Kniehl2009} and MC@NLO~\cite{mcatnlo_hera}. 
}
\label{Tab_NLO_parameters}
\end{center}
\end{table}
\renewcommand{\arraystretch}{1.0} 

\begin{description}
 \item[FMNR:]
The FMNR program~\cite{Frixione,FMNR} is based on an NLO 
calculation in the massive scheme in the collinear approach. 
The resolved and direct processes are calculated 
separately. The program provides weighted parton level events with 
two or three outgoing partons, i.e.~a charm quark pair and possibly one 
additional light parton.
The fragmentation of a charm quark to a \dstar\ meson is treated by 
a downscaling of the three-momentum of the quark  in the charm-anticharm 
rest frame according to the Peterson fragmentation function with a 
parameter value of $\epsilon=0.035$.
The PDF sets HERAPDF1.0\footnote{The HERAPDF1.0 set 
was determined from inclusive deep-inelastic scattering data 
from the H1 and ZEUS experiments in
the GMVFNS. It has been checked that the difference to a PDF set
determined in the massive scheme, CTEQ5F3~\cite{cteq5f3}, is 
significantly smaller than the effect of the
variations considered for the systematic uncertainty of the FMNR 
predictions.}~\cite{Herapdf} for the proton
and GRV-G HO~\cite{grvlo} for the photon are used. For the strong coupling,
the five-flavour QCD scale 
$\Lambda^{(5)}_{QCD}$ is set to $0.2626\ {\rm GeV}$. The charm mass is set to
$m_{c} = 1.5\ {\rm GeV}$ and varied by $\pm 0.2\ {\rm GeV}$ 
for an uncertainty estimate. 
This variation covers the central value for the pole mass
of the charm quark~\cite{pdg10}.
The renormalisation and factorisation scale are set to 
$\mu_{r} = m_{T}$ and $\mu_{f} = 2 \cdot m_{T}$ with $m_{T}$ being
the transverse mass
defined as $m_T^2 = m_c^2+(p_{T,c}^2+p_{T,\bar{c}}^2)/2$,
with $p_{T,c}$ and $p_{T,\bar{c}}$ denoting the transverse momenta of 
the charm and anticharm quark, respectively.
In order to estimate the uncertainties related to missing higher orders, the 
renormalisation and factorisation scales are varied by a factor $2$ up and 
down. Each variation is done independently, leading to in total $6$ variations.
The resulting uncertainties are added in quadrature separately for positive
and negative deviations to obtain the total uncertainties.

\item[GMVFNS:]
A next to leading order cross section prediction for direct and resolved
contributions to the cross section has been provided in the 
GMVFNS~\cite{Kramer, Kniehl2009}.  The 
transition from the charm quark to the \dstar\ meson is given by the 
KKKS fragmentation function which takes 
DGLAP evolution and finite-mass effects into account~\cite{KKK06}.
The parton contents of the proton and of the photon are described by
the PDF sets HERAPDF1.0~\cite{Herapdf} and AFG04~\cite{afg04}, respectively.
The charm mass is set to $m_{c} = 1.5\ {\rm GeV}$, and the
renormalisation and factorisation scales are chosen to be 
$\mu_{r} = \mu_{f} = m_{T}$. The uncertainties related to 
missing higher orders are estimated by varying
the renormalisation scale,
the factorisation scale for the initial state and the factorisation scale
for the final state independently by a factor $2$ up and down 
while satisfying the condition that the ratio of any 
of the two scales is $1/2$, $1$ or $2$. This leads to $14$ independent
variations.
The maximum and minimum values found by this procedure are used to 
determine the systematic uncertainty~\cite{Kniehl2009}.

\item[MC@NLO:] \label{mcatnlo-description}
In the MC@NLO framework~\cite{mcatnlo}, predictions for heavy
flavour production at HERA~\cite{mcatnlo_hera} are provided which
combine an NLO calculation in the massive approach with
parton showers and hadronisation. 
The direct and resolved part of the cross section are
calculated separately. MC@NLO uses parton showers with angular ordering to
simulate higher order contributions
and the cluster fragmentation as implemented in HERWIG~\cite{herwig}. 
A factor of $1.34$ is applied to the MC@NLO predictions in order
to correct the $c \rightarrow \dstar$ branching
fraction in HERWIG to the experimental value~\cite{gladilin}.
The PDF sets HERAPDF1.0~\cite{Herapdf} for the proton
and GRV-G HO~\cite{grvlo} for the photon are used.
For an estimation of the uncertainty, the charm mass and the
renomalisation and factorisation scales are varied separately, and 
the resulting uncertainties 
are added in quadrature.

\end{description}

\section{H1 Detector}
A detailed description of the H1 detector can be found
elsewhere~\cite{h1detector}. 
Only the components  essential to the present analysis are described here. 

The origin of the H1 coordinate system is the nominal $ep$ interaction point.
The positive $z$-axis (forward direction) is defined by the direction of the 
proton beam. Transverse momenta are measured in the $x$--$y$ plane.  
Polar~($\theta$) and~azimuthal~($\varphi$) angles are measured with respect to 
this reference system. The pseudorapidity is defined as 
$\eta = - \ln{\tan(\theta/2)}$.

Charged particles are measured within the central tracking detector (CTD) 
in the pseudorapidity range $-1.74 < \eta < 1.74$. The CTD comprises two 
large cylindrical jet chambers (inner CJC1 and outer CJC2) and the silicon 
vertex detector~\cite{cst}. The CJCs are separated by a drift chamber which 
improves the $z$ coordinate reconstruction. A multiwire proportional 
chamber mainly used for triggering~\cite{mwpc} is situated inside the 
CJC1. These detectors are arranged concentrically around the interaction 
region in a solenoidal magnetic field of \mbox{$1.16\ {\rm T}$}. The 
trajectories of the charged particles are measured with a transverse 
momentum resolution of $\sigma(p_T)/p_T \approx 0.5\% \, p_T/{\rm GeV} 
\oplus 1.5\%$~\cite{ctdresolution}. The CJCs also provide a measurement 
of the specific ionisation 
energy loss ${\rm d}E/{\rm d}x$ of charged particles. 
The interaction vertex is reconstructed from CTD tracks.
The CTD also provides trigger information based on track segments 
measured in the CJCs~\cite{ftt}. At the first two levels of this
fast track trigger (FTT) tracks are reconstructed online from the 
track segments in the CJCs. At the third level of the FTT invariant 
masses of combinations of tracks are calculated~\cite{nik,andy}.

Charged and neutral particles are measured with the liquid argon (LAr) 
calorimeter, which surrounds the tracking chambers. It covers the range 
$-1.5 < \eta < 3.4$ with full azimuthal acceptance. Electromagnetic shower 
energies are measured with a precision of 
$\sigma(E)/E = 12\% / \sqrt{E/{\rm GeV}} \oplus 1\%$ and hadronic energies 
with $\sigma(E)/E = 50\% / \sqrt{E/{\rm GeV}} \oplus 2\%$, as determined 
in test beam measurements~\cite{h1testbeam}.
A lead-scintillating fibre calorimeter (SpaCal)~\cite{spacal} covering 
the backward region $-4.0<\eta<-1.4$ completes the measurement of charged 
and neutral particles. For electrons a relative energy resolution of 
$\sigma(E)/E = 7\% / \sqrt{E/{\rm GeV}} \oplus 1\%$ is reached, as determined 
in test beam measurements~\cite{spacaltestbeam}. 
The hadronic final state is reconstructed using an energy flow algorithm
which combines charged particles measured in the CTD with information from
the SpaCal and LAr calorimeters~\cite{hadroo2}.

The luminosity determination is based on the measurement of the 
Bethe-Heitler process $ep \rightarrow ep\gamma$ where the photon is 
detected in a calorimeter located at $z= -104\ {\rm m}$ downstream of the 
interaction region in the electron beam direction.

\section{Event Selection and Reconstruction}
The data sample was recorded in the years 2006 and 2007, when 
electrons with an energy of $27.6\ {\rm GeV}$ were collided with protons with 
$920\ {\rm GeV}$. 

Photoproduction events are selected by requiring that no isolated high 
energy electromagnetic cluster, consistent with a signal from a scattered 
electron, is detected in the calorimeters. This limits the photon
virtuality to $Q^2 < 2\ {\rm GeV}^2$.

\subsection{\boldmath Inclusive \dstar\ Sample}

The triggering of the events relies on the reconstruction of the final 
state particles originating from the \dstar\ decay. For this purpose all
three levels of the FTT are used. At the first level, where tracks
are reconstructed only in the transverse plane, 
the selection criteria are based on track multiplicities
above certain transverse momentum thresholds. These conditions
are refined on the second level, and on the third level invariant masses 
and charge combinations consistent with the decay channel
$\dstarpm\rightarrow D^{0}\pi^{\pm}_{slow} \rightarrow 
K^{\mp}\pi^{\pm}\pi^{\pm}_{slow}$ are required~\cite{andy}.
Three trigger conditions with different thresholds for the 
transverse momentum of the \dstar\ candidate are used. The analysis 
is therefore performed in three separate $p_T(\dstar)$ regions 
corresponding to the different luminosities:
${\cal L}=30.7\ {\rm pb}^{-1}$ for $1.8 \le p_T(\dstar)< 2.5\ {\rm GeV}$, 
${\cal L}=68.2\ {\rm pb}^{-1}$ for $2.5 \le p_T(\dstar)< 4.5\ {\rm GeV}$, and 
${\cal L}=93.4\ {\rm pb}^{-1}$ for $p_T(\dstar)\ge 4.5\ {\rm GeV}$.
The requirement that all decay particles have to be in the acceptance
of the CJC limits the analysis to central rapidities for the 
\dstar\ meson $|\eta(\dstar)|<1.5$ and photon-proton centre-of-mass 
energies in the range $100 < W_{\gamma p}< 285\ {\rm GeV}$.

The $\gamma p$ centre-of-mass energy is reconstructed using the 
Jacquet-Blondel method~\cite{jacquetblondel}: 
$W_{\gamma p} = \sqrt{y_{JB} \ s}$ with 
$y_{JB} = \sum_{HFS} (E-p_z)_i / (2 \ E_e)$, where $s$ and $E_e$ denote 
the square of the $ep$ centre-of-mass energy and the energy of the incoming 
electron, respectively, and the sum $\sum_{HFS}$ runs over the energy $E$ 
and the longitudinal momentum $p_z$ of all final state particles.
The \dstar\ inelasticity \zDs, which corresponds to the fraction of 
photon energy transferred to the \dstar\ meson in the proton rest frame,
is defined by $\zDs = P \cdot p(\dstar)/(P\cdot q)$, 
with $P$, $p(\dstar)$ and $q$ denoting the four-momenta of the incoming 
proton, the \dstar\ meson and the exchanged photon, respectively. It is 
reconstructed as $\zDs= (E-p_{z})_{\dstar}/(2 \ y_{JB} \ E_{e})$.
The inelasticity distribution is sensitive to the kinematics of the 
production mechanism and to the $c\rightarrow \dstar$ fragmentation function. 

The \dstar\ meson is detected via the decay channel
$\dstarpm\rightarrow D^{0}\pi^{\pm}_{slow} 
\rightarrow K^{\mp}\pi^{\pm}\pi^{\pm}_{slow}$ with a branching fraction of 
${\cal BR}= 2.63 \pm 0.04\%$~\cite{pdg10}. The tracks of the decay 
particles are reconstructed using the CTD information. The invariant mass 
of the $K^{\mp}\pi^{\pm}$ system is required to be consistent with the nominal 
$D^{0}$ mass~\cite{pdg10} within $\pm 80\ {\rm MeV}$. 
The signal to background ratio is improved by applying a loose particle 
identification criterion to the kaon candidates based on the measurement 
of the specific energy loss, ${\rm d}E/{\rm d}x$, in the CTD. 
In addition the background is reduced by 
a cut on the fraction of the transverse momentum carried by the \dstar\ 
with respect to the scalar sum of transverse energies of the hadronic 
final state, excluding the forward region ($\theta < 10^\circ$).
This fraction is required to be 
$p_{T}(\dstar)/(\sum_{HFS}^{\theta > 10^\circ} E_{T,i})>0.1$. 
This criterion accounts for the 
harder fragmentation of charm compared to light flavours.

\begin{figure}[ht]
     \centering
\unitlength1.0cm
\begin{picture}(16,18)
\put(1.5,9.5){\includegraphics[width=13cm]{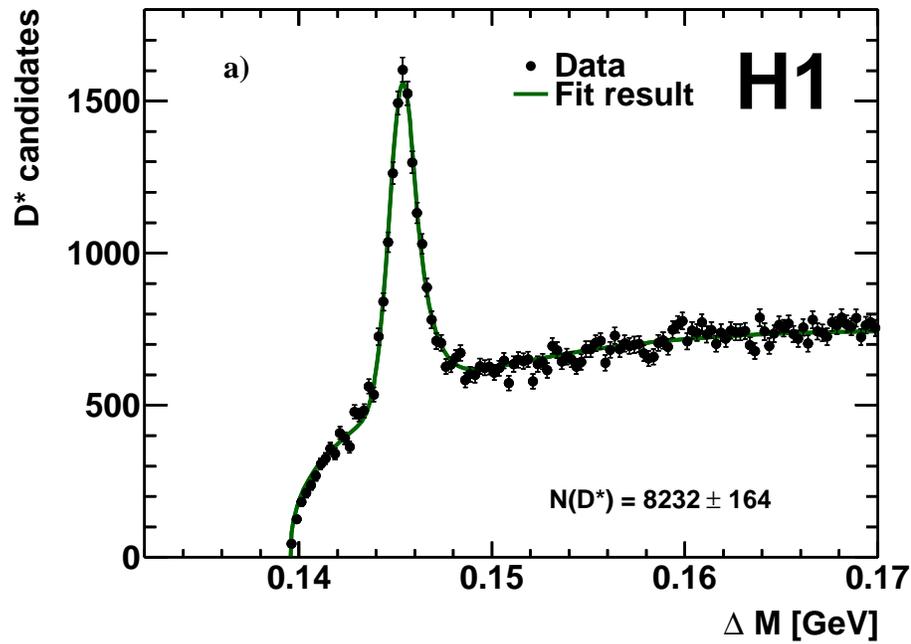}}
\put(1.5,0){\includegraphics[width=13cm]{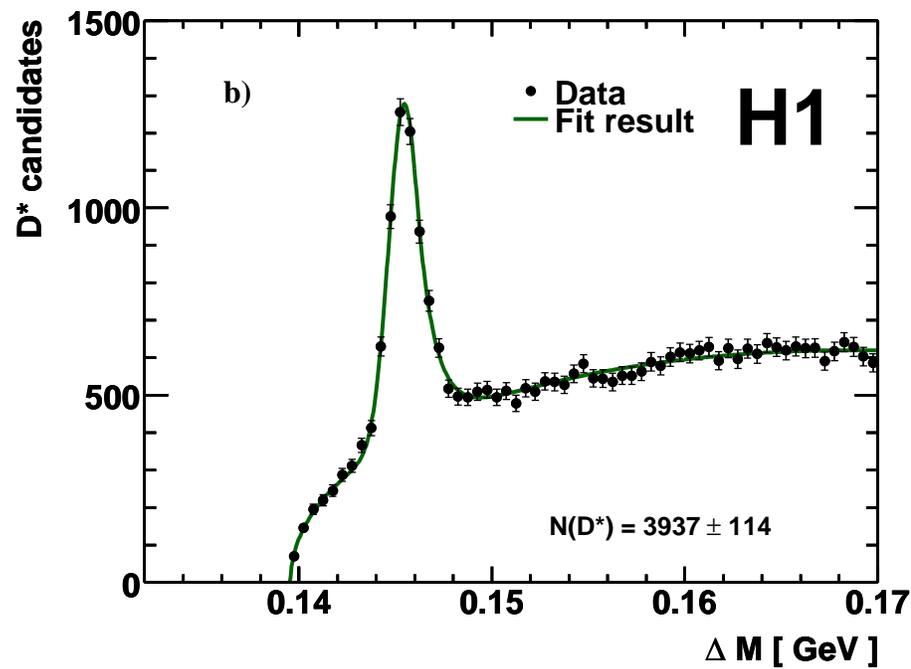}}
\put(4.5,17.){\bf a)}
\put(4.5,7.5){\bf b)}
\end{picture}
    \caption{Distribution of $\Delta M$ for \dstar\ candidates a) in the
    inclusive \dstar\ sample and b) in the \dstar\ tagged dijet sample. 
    The fit function is also shown.}
    \label{signal}
\end{figure}

The \dstarpm\ candidates are selected using the mass difference 
method~\cite{deltammethod}. In figure~\ref{signal}a) the distribution 
of the mass difference $\Delta M = m(K\pi \pi_{slow})-m(K\pi)$ of the 
final \dstar\ candidates is shown. A clear peak is observed around 
the nominal value of $\Delta M = 145.4 \ {\rm MeV}$~\cite{pdg10}.

The wrong charge combinations, defined as $K^{\pm}\pi^\pm \pi_{slow}^{\mp}$ 
with $K^\pm \pi^\pm$ pairs in the accepted $D^0$ mass range, are used to 
constrain the shape of the combinatorial background in the signal region. 
The number of reconstructed \dstar\ mesons $N(\dstar)$ is extracted in 
each analysis bin by a log-likelihood fit simultaneously to the right charge 
and the wrong charge $\Delta M$ distribution. For the signal which has 
a tail towards larger $\Delta M$ values the asymmetric Crystal Ball 
function~\cite{Gaiser} is used. The shape of the background is 
parametrised with the Granet function~\cite{Granet}.
The fit is performed in the RooFit framework~\cite{Verkerke}. 
The fit to the inclusive data sample yields $8232 \pm 164$ \dstar\ mesons.
To improve the convergence of the fit in each analysis bin,
the parameters describing the asymmetry of the Crystal Ball function 
are fixed to the values found by the fit to the complete data set. The 
width of the peak varies 
in dependence on the \dstar\ kinematics and is therefore left free.
More details can be found in~\cite{thesis_eva}.

\subsection{\boldmath \dstarDj\ Sample}

For the selection of the \dstar\ meson in the \dstarDj\ sample, the
requirements are the same as for the inclusive \dstar\ sample, except that
the requirement on the specific energy loss ${\rm d}E/{\rm d}x$ is removed, 
and the cut on $p_T(\dstar)$ is increased to $2.1\ {\rm  GeV}$ 
because of large backgrounds at small transverse momenta.

Jets are defined by the inclusive
$k_t$-algorithm~\cite{jetKT93} in the  energy re\-com\-bina\-tion
scheme with jet size
$\Delta R = \sqrt{(\Delta \eta)^2 + (\Delta \varphi)^2} = 1$
where $\Delta \varphi$ is expressed in radians.
The jet algorithm is applied in the laboratory frame to all reconstructed 
particles of the hadronic final state.
To prevent the decay particles of the \dstar\ candidate from being 
attributed to different jets,
the \dstar\ candidate is used as a single particle in the jet algorithm,
replacing its decay products.
In this way the jet containing the \dstar\ meson (\dstarjet) is unambiguously 
defined for each \dstar\ candidate.
In events which contain more than one \dstar\ candidate, the jet algorithm 
is run separately for each candidate, and all candidates for which the
dijet selection criteria are fulfilled enter the $\Delta M$ distribution. 
The pseudorapidity of the \dstarjet\ is restricted to the same range as is used
for the \dstar\ meson, $|\eta(\dstarjet)| < 1.5$.
In addition to the \dstarjet\ a second jet is required. Both jets have to
satisfy $\ptj > 3.5\ {\rm GeV}$. 
If there is more than one jet that does not contain the \dstar\ meson, 
the one with the highest \ptj\ is chosen as the \otherj.
The pseudorapidity of the \otherj\ has to be in the range
$-1.5 < \eta(\otherj) < 2.9$. 
The invariant mass $M_{jj}$ of the \dstarjet\ 
and the \otherj is required to satisfy $M_{jj} > 6\ {\rm GeV} $ in order 
to select jets from the partons originating from the hard interaction.
More details on the selection of the \dstarDj\ sample
can be found in~\cite{thesis_zlatka}.

The number of \dstarDj\ s is extracted from the $\Delta M$ distribution 
of the \dstar\ candidates with the same procedure as used  for the 
inclusive \dstar\ measurement.
The $\Delta M$ distribution for the selected events in the dijet sample 
is shown in figure~\ref{signal}b). The fit yields a signal of
$3937 \pm 114$ \dstar\ mesons.


The kinematic 
range of the inclusive \dstar\ measurement and of the \dstarDj\ 
measurement are summarised in table~\ref{tab:kinrange}.

\renewcommand{\arraystretch}{1.15}
\begin{table}[h]
\centering
\begin{tabular}{|l|l|}
\hline
\multicolumn{2}{|c|}{\bf \boldmath inclusive \dstar\ meson and \dstarDj\ production} \\
\hline
 Photon virtuality & $Q^{2} < 2\ {\rm GeV}^{2}$\\
 $\gamma p$ centre-of-mass energy & $100 < W_{\gamma p}< 285\ {\rm  GeV}$ \\ 
 Pseudorapidity of \dstarpm & $|\eta(\dstar)| < 1.5$ \\
 \hline \hline
\multicolumn{2}{|c|}{\bf \boldmath inclusive \dstar\ meson production} \\
\hline
 Transverse momentum of \dstarpm & $p_{T}(\dstar) > 1.8\ {\rm GeV}$ \\
 \hline \hline
\multicolumn{2}{|c|}{\bf \boldmath \dstarDj\ production} \\
\hline
  Transverse momentum of \dstarpm & $p_{T}(\dstar) > 2.1\ {\rm GeV}$ \\
Transverse momentum of \dstarjet & $p_{T}(\dstarjet) > 3.5\ {\rm GeV}$ \\
  Pseudorapidity of \dstarjet & $|\eta(\dstarjet)| < 1.5$ \\
  Transverse momentum of \otherj & $p_{T}(\otherj) > 3.5\ {\rm GeV}$ \\
  Pseudorapidity of \otherj  & $-1.5 < \eta(\otherj) < 2.9$ \\
  Dijet invariant mass $M_{jj}$ &  $M_{jj} > 6\ {\rm GeV} $\\
  \hline
\end{tabular}
\caption{Definition of the kinematic range of the measurements. 
\label{tab:kinrange}
}
\end{table}
\renewcommand{\arraystretch}{1.0}

\section{Cross Section Determination and Systematic Errors}

The bin averaged visible differential cross section with respect to a 
variable $Y$ (with bin width $\Delta Y$) is calculated according to
\begin{equation}
\frac{{\rm d}\sigma_{vis}(ep\rightarrow e\ \dstar + X)}{{\rm d}Y} = 
\frac{N(\dstar)(1-r)}{\Delta Y \cdot {\cal L}\cdot {\cal BR} \cdot \epsilon } 
\label{eq:crossec}
\end{equation}     
where ${\cal L}$ is the integrated luminosity, ${\cal BR}$ is the 
branching ratio of the analysed decay chain 
$\dstarpm\rightarrow D^{0}\pi^{\pm}_{slow} 
\rightarrow K^{\mp}\pi^{\pm}\pi^{\pm}_{slow}$
and $(1-r)$ a correction factor 
to account for reflections from other $D^0$ decays. The efficiency $\epsilon$
includes the detector acceptance, trigger and reconstruction efficiencies
and migrations between bins. 
The contributions of \dstar\ mesons originating from beauty
production and from gluon splitting from light flavour production is not
subtracted. It is estimated from MC predictions to be below $2 \%$.

The systematic uncertainties are determined in each bin separately
and are summarised in table \ref{tab:sysError} for the total cross section.
They are divided into uncertainties which are considered to be
uncorrelated between the bins and uncertainties which change the
cross section normalisation in all bins.
The numbers 
for the uncertainties listed below are given in per cent
of the cross section values.

The following uncorrelated systematic uncertainty sources are considered:
\begin{description}
\item[Trigger Efficiency:]
The simulation of the FTT is verified by a comparison to data in a 
sample of \dstar\ mesons in deep-inelastic scattering triggered by
the scattered electron. For the total inclusive \dstar\ sample the 
efficiency agrees within a relative uncertainty of $7.5\%$. 
This is one of the dominant systematic uncertainties. For the
\dstarDj\ sample the trigger efficiency is higher, leading to a smaller
uncertainty of $3.1\%$ for the total cross section.
\item[Signal Extraction:]
For the determination of the uncertainty of the signal fit, different 
parameterisations for the signal and background functions are used. 
The resulting uncertainty amounts to $1.5\%$.
\item[\boldmath $D^0$ mass cut:] 
The loss of \dstar\ mesons due to the $D^0$ mass cut is compared between
data and simulation as a function of the \dstar\ transverse momentum, assuming
a Gaussian resolution for the $D^0$ mass reconstruction. 
They agree within $2\%$, which is assigned as uncertainty.
\item[Reflections:]
The amount of reflections $r$ from decay modes of the $D^0$ meson other 
than $D^0 \rightarrow K^\mp \pi^{\pm}$ amounts to $3.8\%$ in the 
simulation~\cite{h1dstardis}. It is independent of kinematic quantities 
within $1\%$, which is used as systematic uncertainty.
\item[Background from deep inelastic scattering:]
The background originating from deep inelastic scattering events is 
estimated with the RAPGAP~\cite{rapgap} MC generator. It is found to 
be below $1\%$, which is not subtracted but treated as an uncertainty.
\item[\boldmath ${\rm d}E/{\rm d}x$ cut:]
The efficiency of the cut on the ${\rm d}E/{\rm d}x$ likelihood of the 
kaon candidate is studied for data and MC simulation in bins of the 
transverse momentum of the \dstar\ meson. The relative difference of $1.5\%$ 
is corrected for in the MC sample.
An uncertainty of $0.5\%$ is assigned, covering the possible 
$p_{T}(\dstar)$ dependence of this correction. 
\item[Hadronic energy scale:]
The energy scale of the hadronic final state has an uncertainty of $2\%$
leading to an uncertainty of the cross section of $0.6\%$ in the inclusive 
\dstar\ sample and of $2.0\%$ in the \dstarDj\ sample.
\item[Model:]
For the determination of the cross section the \PYTHIA\ and \CASCADE\ 
predictions are reweighted to describe the data distributions where
necessary. For the
correction of the data the efficiency from the \PYTHIA\
MC is used. The difference to the efficiency from \CASCADE\ is taken as
a systematic uncertainty. 
It amounts to $2\%$ ($1.5\%$) for the total inclusive \dstar\ (\dstarDj )
cross section. 
\item[Fragmentation:]
The $\alpha$ parameter of the Kartvelishvili function and the position of 
the $\hat{s}$ threshold are varied within the values given in 
table~\ref{fragpar} resulting in an uncertainty of $2.5\%$ ($2.0\%$) 
for the total inclusive \dstar\ (\dstarDj ) cross section.  
\end{description}

The following normalisation uncertainties are considered:
\begin{description}
\item[Track finding efficiency:]
The systematic uncertainty on the track efficiency of $4.1\%$ per \dstar~meson 
arises from two contributions: (i) The comparison of the track finding efficiency
in data and simulation leads to an uncertainty of $2\%$ for the slow pion track
and $1\%$ for the tracks of the $D^0$ decay particles, and the uncertainty
is assumed to be
correlated between the decay particles; (ii) the efficiency with which
a track can be fitted to the event vertex
leads to a systematic error of $1\%$ per \dstar~meson.
The uncertainty on the track finding efficiency is considered to be half
correlated between the bins of the measurement.
\item[Luminosity:]
The uncertainty on the luminosity measurement for the data sample used 
in this analysis amounts to $5\%$. 
\item[Branching Ratio:]
The uncertainty due to the \dstar\ branching ratio is $1.5\%$~\cite{pdg10}.
\end{description}

All sources of systematic errors are added in quadrature resulting in 
a systematic uncertainty of 
$10.9\%$ ($8.5\%$)
for the total cross section of 
the inclusive \dstar\ (\dstarDj )  production.

\begin{table}[htd]
\begin{center}
\begin{tabular}{|l|r|r|}
\hline
Uncertainty source  &   \dstar & \dstarDj \\  \hline
\multicolumn{3}{|l|}{Uncorrelated uncertainties }    \\  \hline
Trigger efficiency &            	$7.5\%$ & $3.1\%$\\
Signal extraction &             	$1.5\%$ & $1.5\%$ \\
$D^0$ meson mass cut &          	$2.0\%$ & $2.0\%$ \\
Reflections &                   	$1.0\%$ & $1.0\%$ \\
Background from deep-inelastic scattering & $1.0\%$ & $1.0\%$ \\
${\rm d}E/{\rm d}x$ cut &       	$0.5\%$ & $-$ \\  
Hadronic energy scale &         	$0.6\%$ & $2.0\%$ \\  
Model  &             			$2.0\%$ & $1.5\%$ \\
Fragmentation &                 	$2.5\%$ & $2.0\%$ \\ \hline
Track finding efficiency (half) & 	$2.9\%$ & $2.9\%$ \\
Total uncorrelated &			$9.2\%$ & $6.0\%$ \\ \hline \hline
\multicolumn{3}{|l|}{Normalisation uncertainties}   \\  \hline
Track finding efficiency (half) & 	$2.9\%$ & $2.9\%$ \\
Luminosity &                    	$5.0\%$ & $5.0\%$ \\
Branching ratio &               	$1.5\%$ & $1.5\%$ \\ \hline
Total normalisation & 			$6.0\%$ & $6.0\%$ \\ \hline \hline
Total &                         	$10.9\%$ & $8.5\%$ \\  \hline

\end{tabular}
\end{center}
\caption{Summary of all sources of systematic uncertainties and their
effect on the
total \dstar\ and the \dstarDj\ production cross section with the breakdown into sources leading to 
bin-to-bin uncorrelated uncertainties and sources leading to 
normalisation uncertainties.
}
\label{tab:sysError}
\end{table}

\section{\boldmath Results for Inclusive \dstar\ Meson Production}
The total visible cross section for \dstar~meson photoproduction is 
measured to be:
\begin{equation}
\sigma_{vis}(e p \rightarrow e \ \dstar + X)=  
41.1 \pm 0.8~({\rm stat.}) \pm 3.6~({\rm unc. sys.})\pm 2.7~({\rm norm.})~{\rm nb}
\end{equation}
in the kinematic range defined in table~\ref{tab:kinrange}. The corresponding 
predictions from \PYTHIA\ and \CASCADE\ amount to $43.7~{\rm nb}$ and 
 $32.9~{\rm nb}$, respectively. Due to the fact that these predictions are based 
 on leading order matrix elements the uncertainty on the normalisation 
 of the cross sections is large, and is not quantified here.
 The NLO calculations predict 
 $26\,^{+ 13}_{-\ \,8}~{\rm nb}$ for FMNR, 
 $37\,^{+28}_{-14}~{\rm nb}$ for GMVFNS and
 $30\,_{- 7}^{+ 6}$ for MC@NLO.
 
\begin{figure}[ht]
\unitlength1.0cm
\begin{picture}(16,18)
\put(0,6.5){\includegraphics[width=8cm]{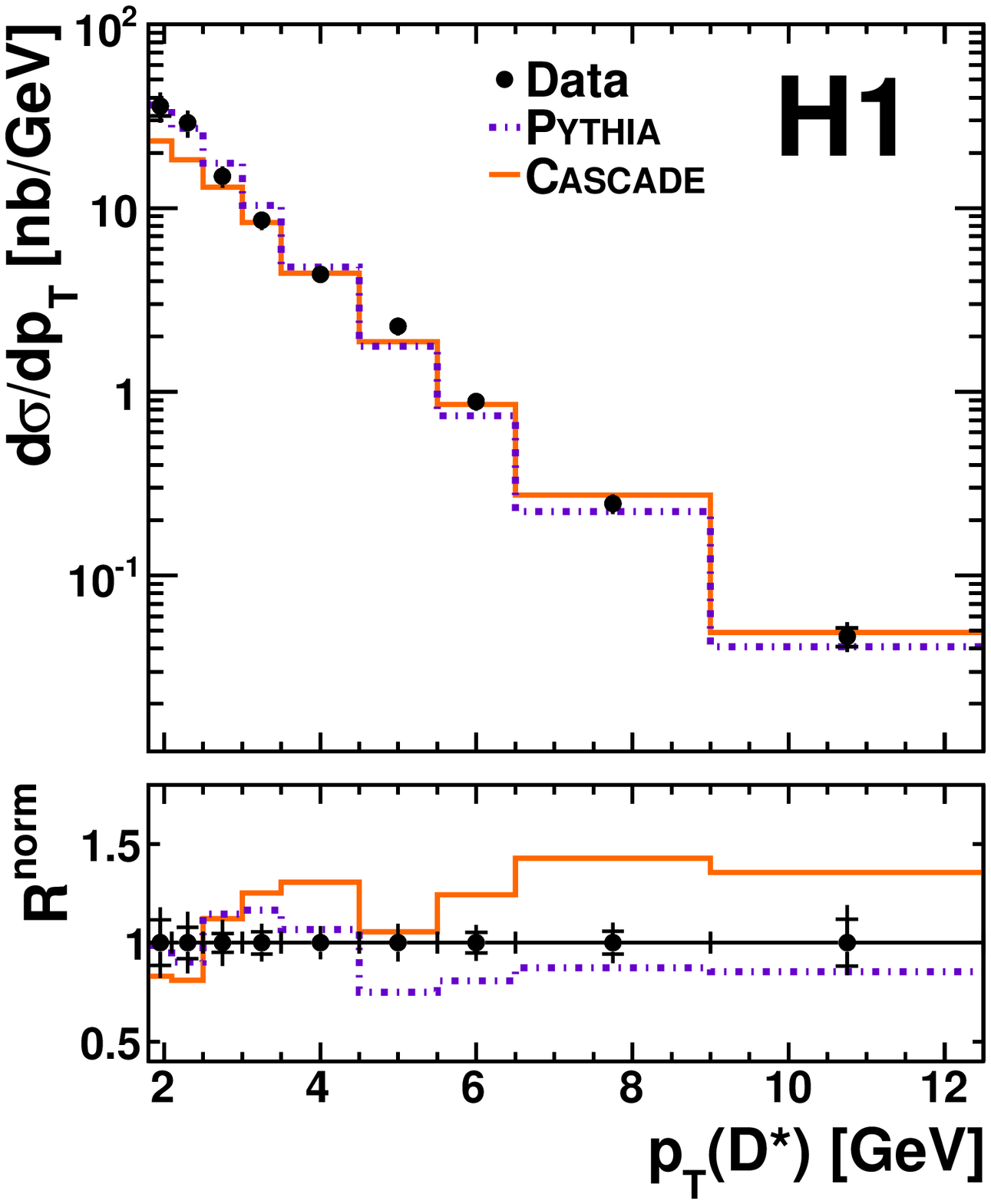}}
\put(8,6.5){\includegraphics[width=8cm]{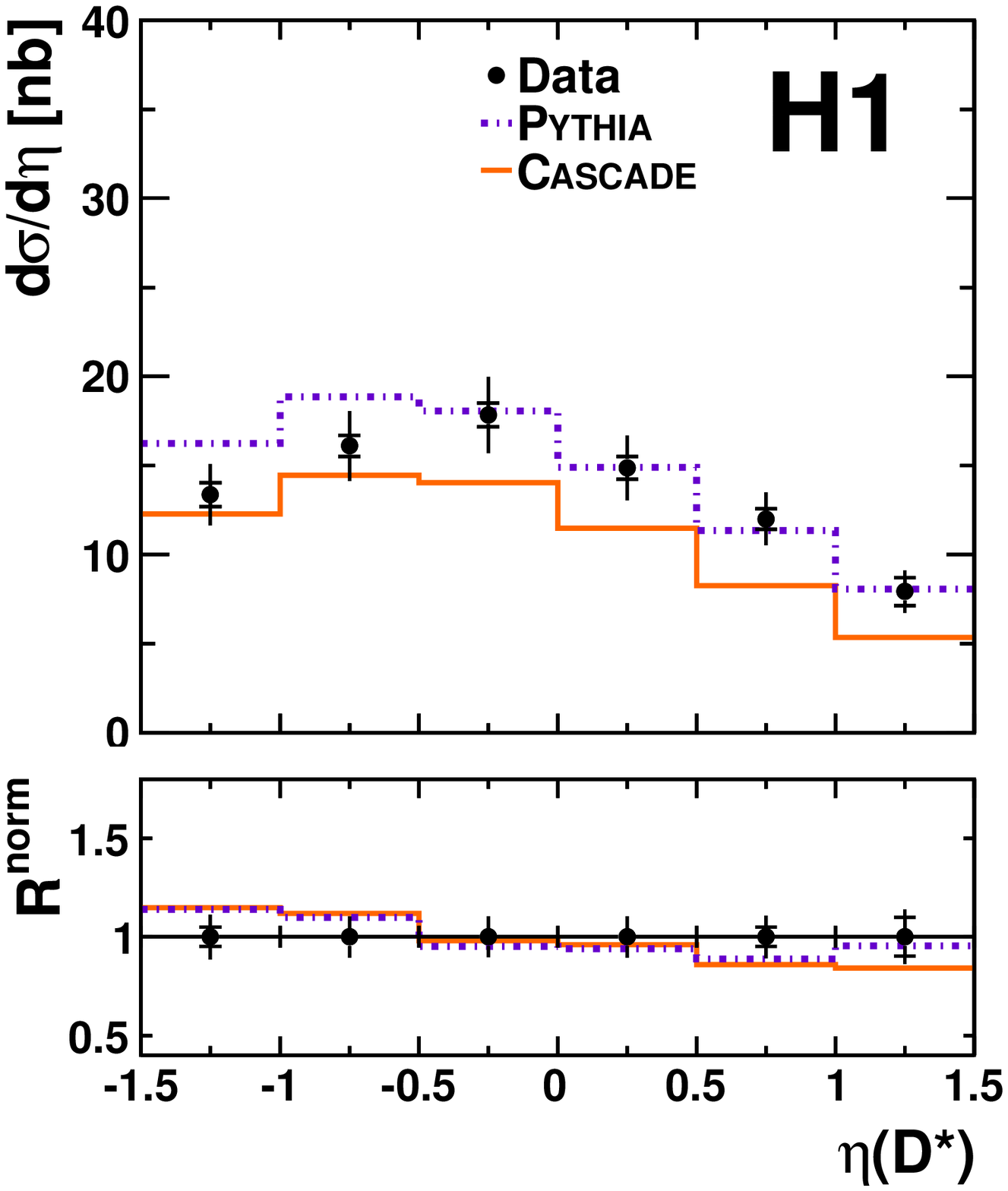}}
\put(0,-3){\includegraphics[width=8cm]{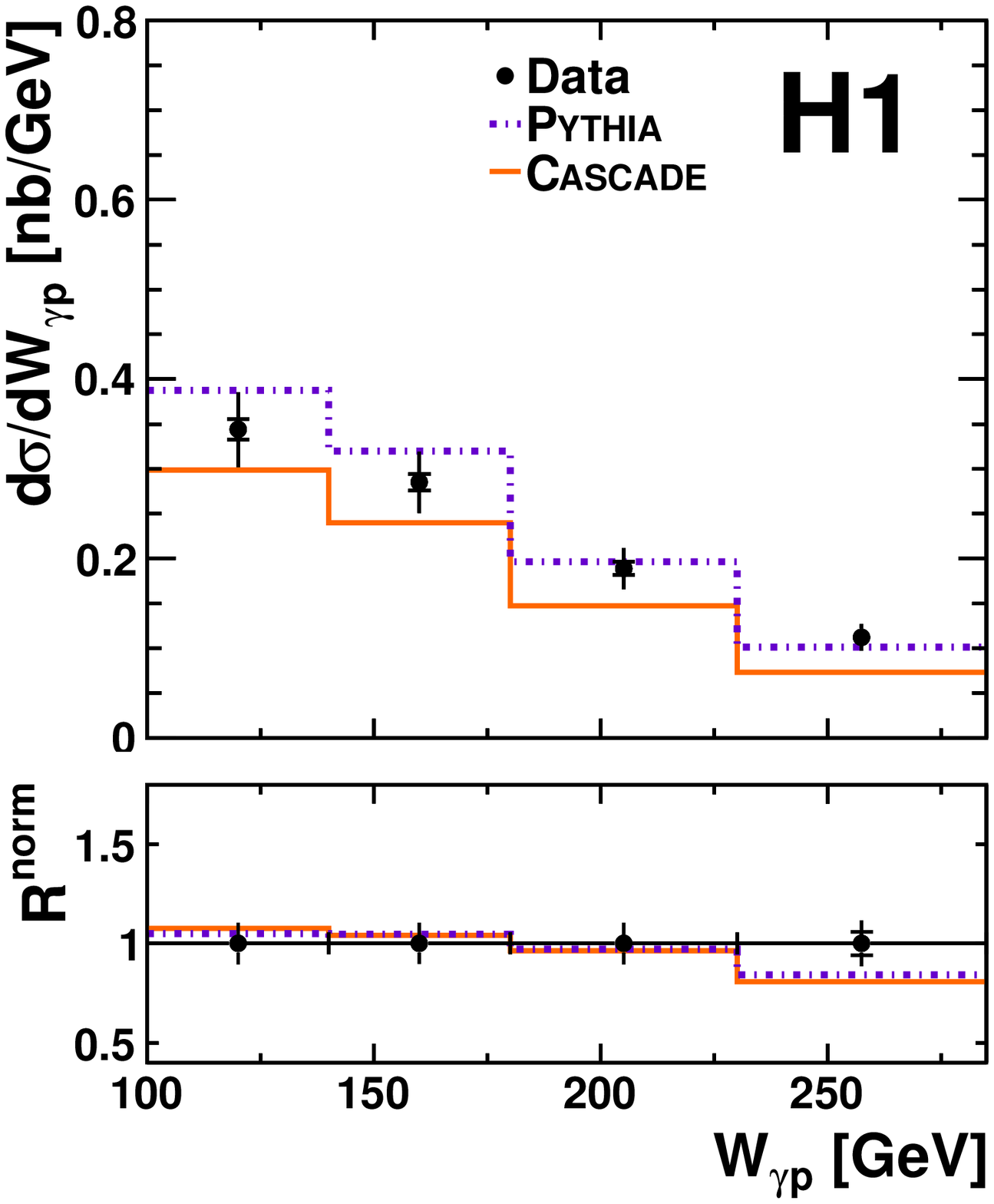}}
\put(8,-3){\includegraphics[width=8cm]{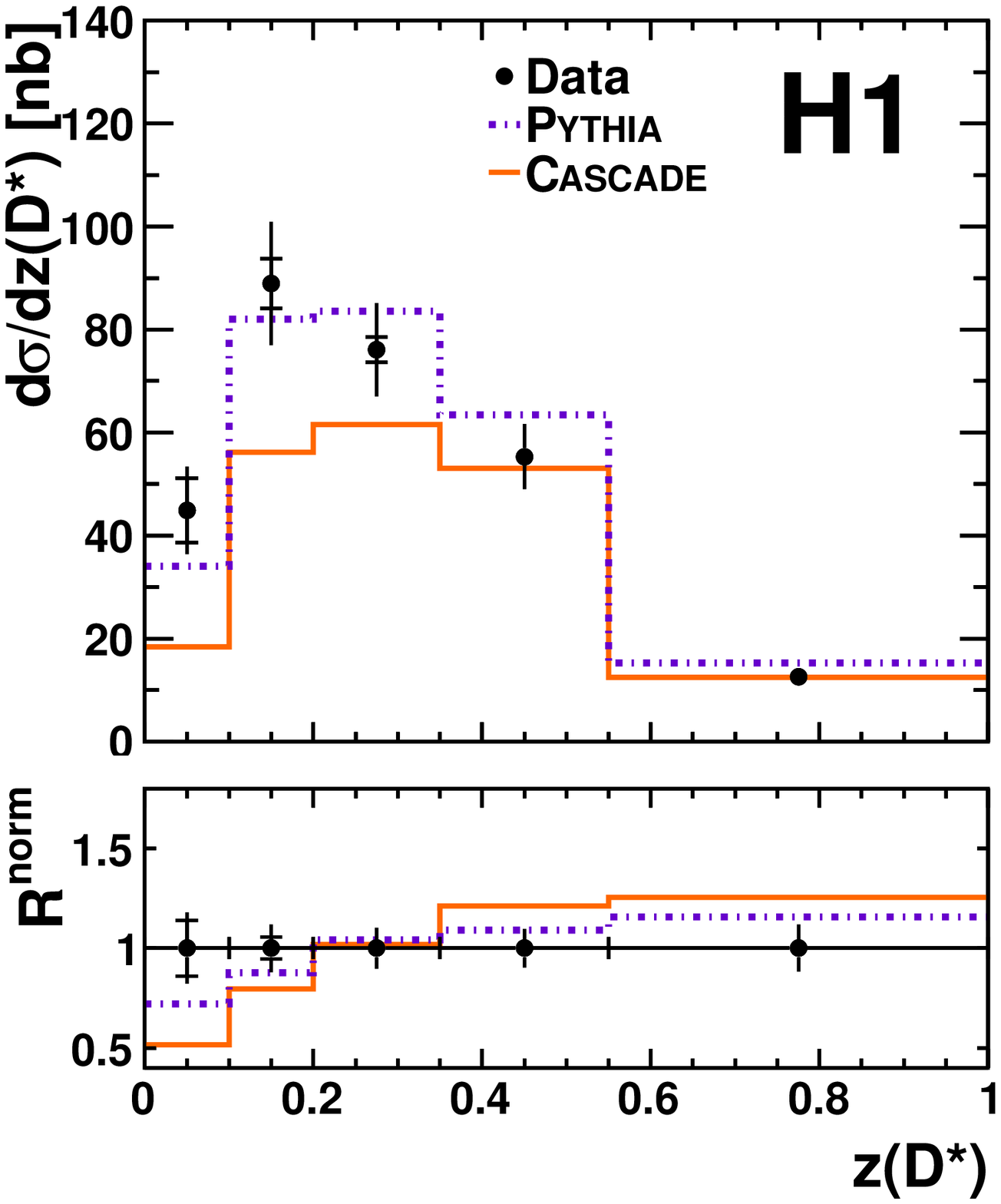}}
\put(1.8,17.5){\bf a)}
\put(9.8,17.5){\bf b)}
\put(1.8,7.5){\bf c)}
\put(9.8,7.5){\bf d)}
\end{picture}
   \caption{Single differential \dstar\ cross section as a function of \ptds, \etads, \Wgp,
   and \zDs\
    compared to \PYTHIA\ and \CASCADE\ predictions. Here and in the following
    figures the inner
    error bar depicts the statistical error and the outer shows the statistical, and 
    uncorrelated systematic and normalisation uncertainty added in 
    quadrature. The normalised ratio \rnorm\
    (see text) is also shown.}
    \label{singlediff}
\end{figure}

\begin{figure}[ht]
\unitlength1.0cm
\begin{picture}(16,18)
\put(0,6.5){\includegraphics[width=8cm]{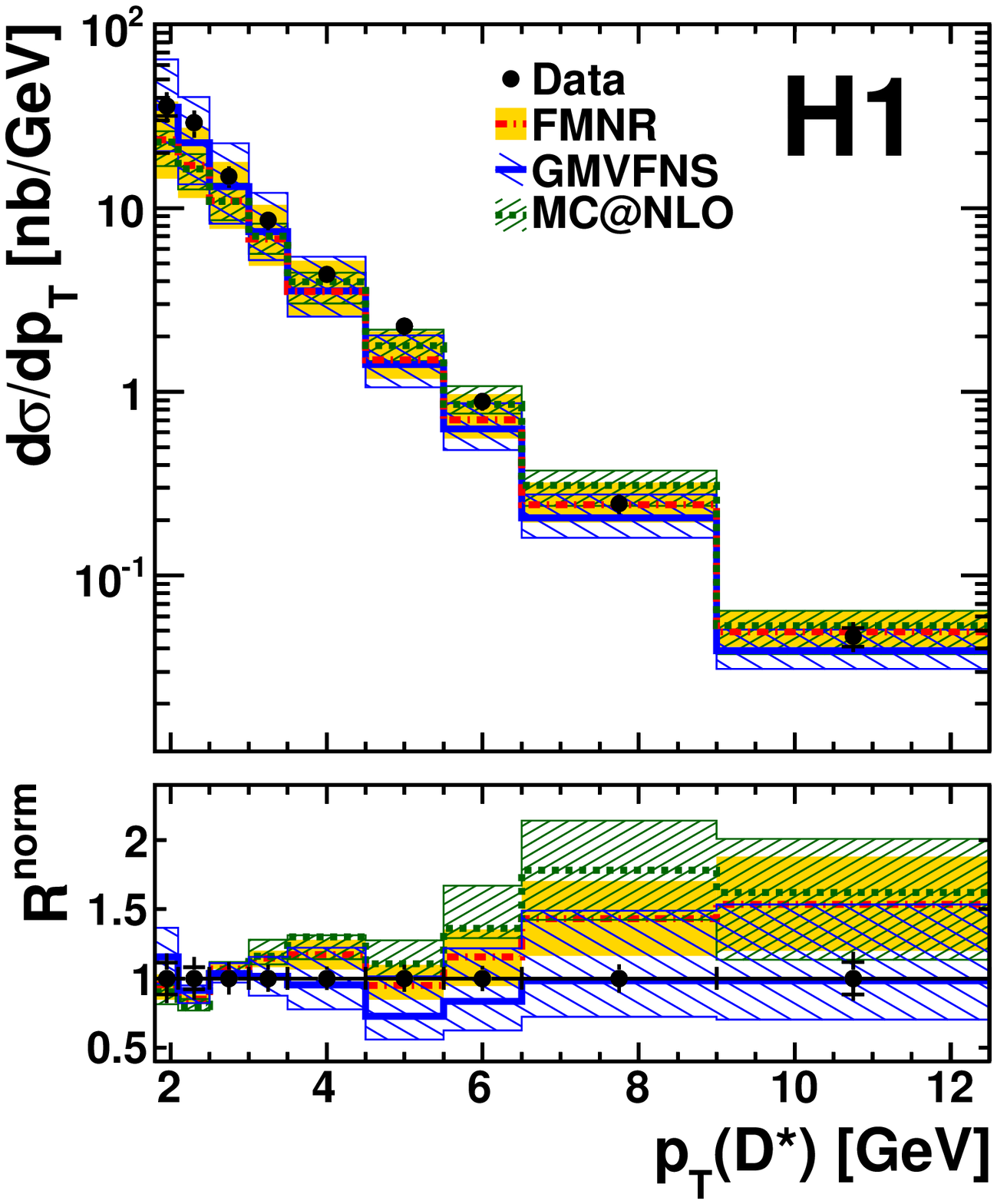}}
\put(8,6.5){\includegraphics[width=8cm]{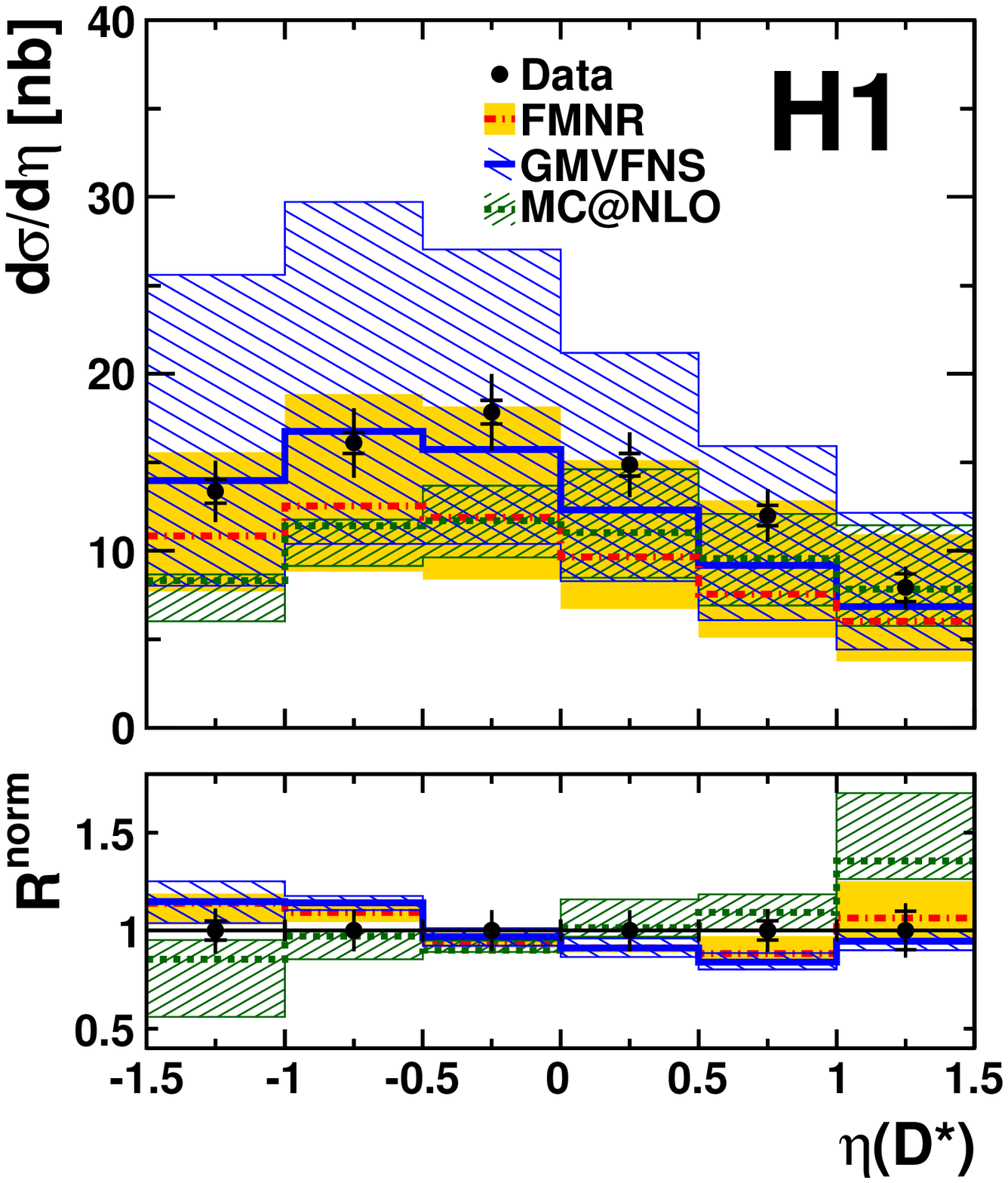}}
\put(0,-3){\includegraphics[width=8cm]{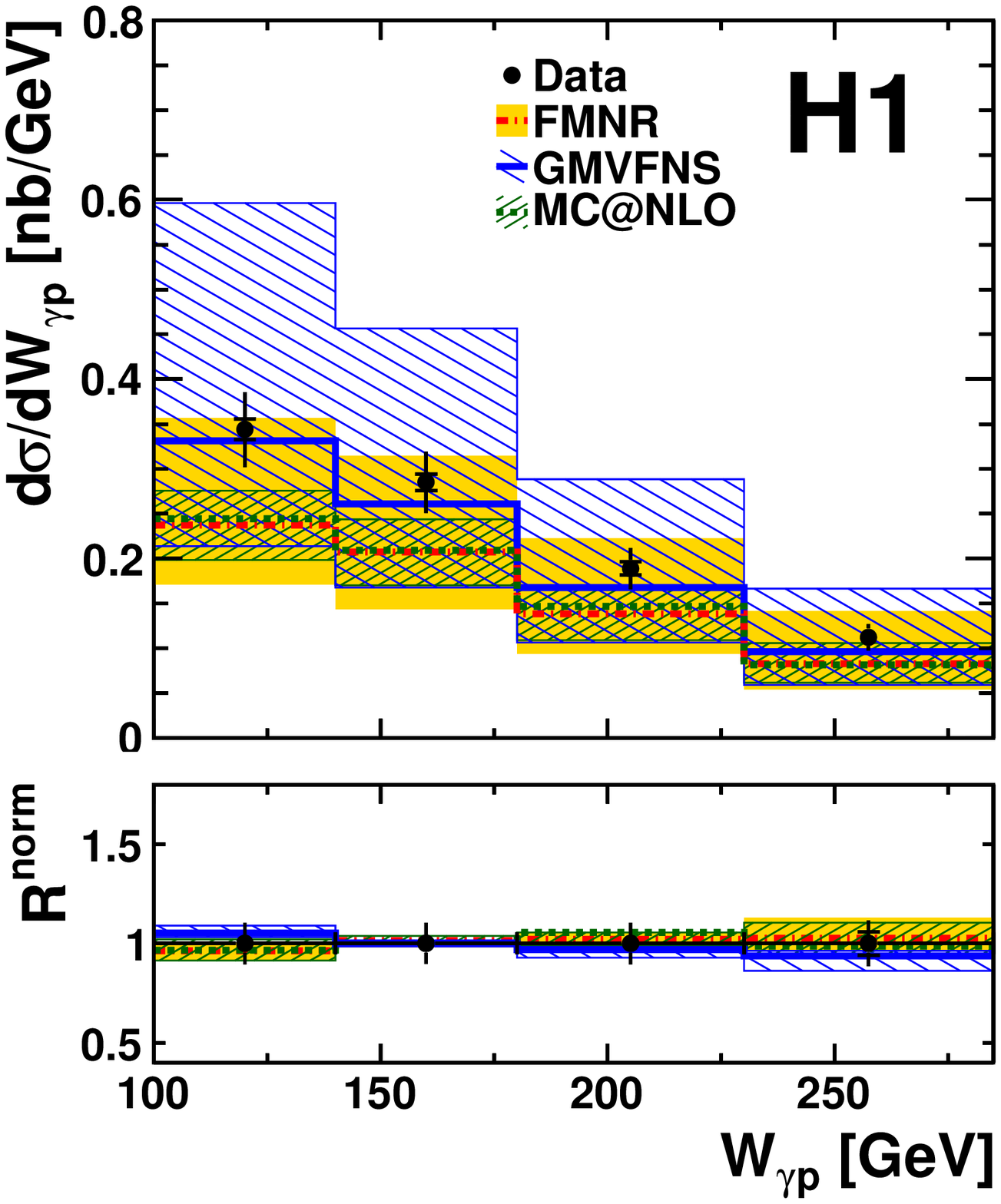}}
\put(8,-3){\includegraphics[width=8cm]{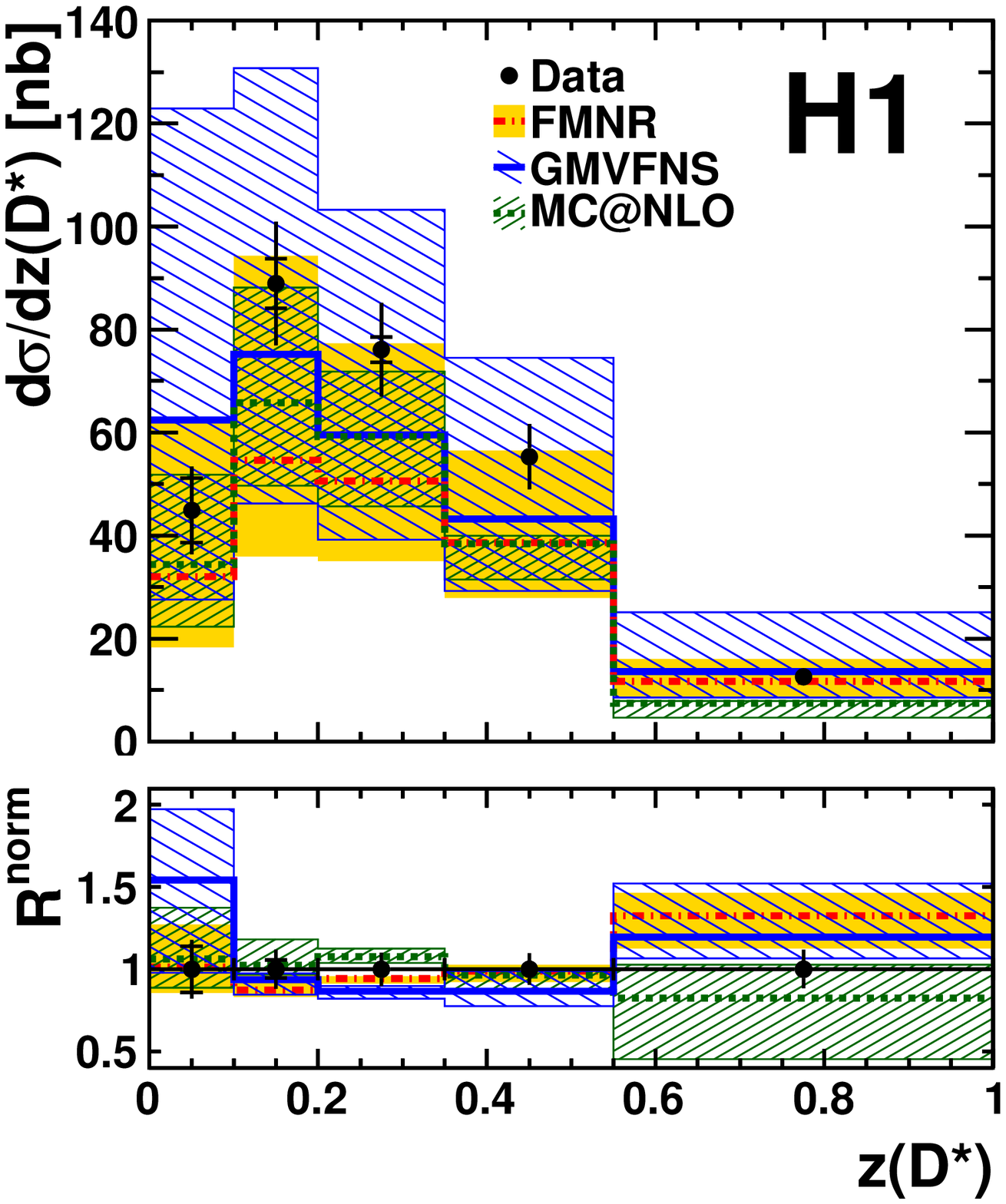}}
\put(1.8,17.5){\bf a)}
\put(9.8,17.5){\bf b)}
\put(1.8,8.){\bf c)}
\put(14.3,6){\bf d)}
\end{picture}
    \caption{Single differential \dstar\ cross section as a function of \ptds, \etads, \Wgp, 
    and \zDs\
    compared to the next-to-leading order predictions of 
    FMNR, GMVFNS and MC@NLO.
    The normalised ratio \rnorm\ (see text) is also shown.}
    \label{singlediff_NLO}
\end{figure}

The measured single differential cross section as a function of the transverse
momentum \ptds\ and the pseudorapidity \etads\ of the \dstar~meson, 
the photon-proton centre-of-mass energy $W_{\gamma p}$ and 
\dstar\ inelasticity \zDs\ are presented in table~\ref{tab:cssd} and 
in figures~\ref{singlediff} 
and~\ref{singlediff_NLO}. The data are compared to \PYTHIA, \CASCADE\ and 
the NLO predictions of FMNR, GMVFNS and MC@NLO.
Since all the predictions have large normalisation uncertainties,
the normalised ratio \rnorm\ of theory to data is shown in order to 
compare the shape of the various predictions to the data. \rnorm\ is 
defined as 
\begin{equation}
 \rnorm = \frac{\dfrac{1}{\sigma_{\rm vis}^{\rm calc}}\cdot 
 \dfrac{{\rm d}\sigma^{\rm calc}}{{\rm d}Y}} 
 {\dfrac{1}{\sigma_{\rm vis}^{\rm data}}\cdot 
 \dfrac{{\rm d}\sigma^{\rm data}}{{\rm d}Y}}
\end{equation} 
where $\sigma_{\rm vis}^{\rm calc}$ ($\sigma_{\rm vis}^{\rm data}$) and 
${{\rm d}\sigma^{\rm calc}}/{{\rm d}Y}$ 
(${{\rm d}\sigma^{\rm data}}/{{\rm d}Y}$) 
are the total and differential cross
section of the model under consideration (of the data), respectively, 
and $Y$ denotes any measured variable. 
In this ratio the normalization uncertainties of the data (luminosity, 
branching ratio and half of the tracking uncertainty) cancel. 
Similarly, uncertainty sources of the NLO predictions altering the 
normalisation only do not affect \rnorm\ since for each variation the 
total and the differential cross section are varied simultanously. 

The single differential cross sections are compared to the predictions
of the LO MC simulations
in figure~\ref{singlediff}. The steep decrease of the cross section 
with increasing transverse momentum \ptds\ is 
reasonably reproduced by \PYTHIA, 
while \CASCADE\ falls slightly slower than the data. 
Both MC simulations describe the shape of the observed \etads\ 
distribution within uncertainties. 
The cross section decreases as a function of the $\gamma p$ centre-of-mass
energy $W_{\gamma p}$, as expected from the photon flux in the equivalent 
photon approximation~\cite{epa}. \CASCADE\ predicts a smaller fraction 
of \dstar~mesons being produced at small inelasticities \zDs, similar to what
has been observed in deep inelastic scattering at HERA~\cite{h1dstardis}.
All distributions are reasonably well described by \PYTHIA.

A comparison of the single differential cross sections to the predictions
of the NLO calculations is shown in figure~\ref{singlediff_NLO}.
For all measured quantities the precision of the measurement presented here 
is much better than the estimated uncertainty of the NLO calculations.
The uncertainty of the NLO predictions is dominated by the variation
of the renormalisation scale $\mu_r$, which has a large effect on the
absolute cross section, while the differences in the shapes tend to be 
smaller. Within these large theoretical uncertainties, both the FMNR 
and GMVFNS predictions agree 
with the measured cross section as a function of \ptds, while the
MC@NLO underestimates the data at small \ptds. The \ptds\ shape 
is best described by the GMVFNS calculation, while FMNR and MC@NLO 
predict a harder spectrum than observed in data as can be seen 
in the ratio \rnorm.
The underestimation of the low \ptds\ region by the central FMNR and MC@NLO 
predictions results 
in a low normalisation in the other distributions. 
The shape of the \etads\ distribution is reasonably well described by all 
NLO calculations. 
All three NLO calculations give a rather precise prediction of the shape 
of the $W_{\gamma p}$ distribution, which describes the measurement.
Given the large uncertainties the predictions for the \zDs\ distribution 
agree with the data, although when using the central parameter
settings for the calculations they differ in shape with respect to data.

Previous H1 and ZEUS analyses of \dstar\ meson 
photoproduction~\cite{zeusdstar98,h1dstar06}, albeit in 
different kinematic ranges in the photon virtuality $Q^2$ and the 
photon-proton centre-of-mass energy $W_{\gamma p}$, lead to similar 
conclusions: while all predictions 
give a good description of the $W_{\gamma p}$ distribution, differences 
between data and theoretical predictions are observed for variables sensitive 
to the quantities of the outgoing charm quark.

\begin{figure}[hp]
     \centering
\unitlength1.0cm
\begin{picture}(14,14)
\put(0,0){\includegraphics[width=14cm]{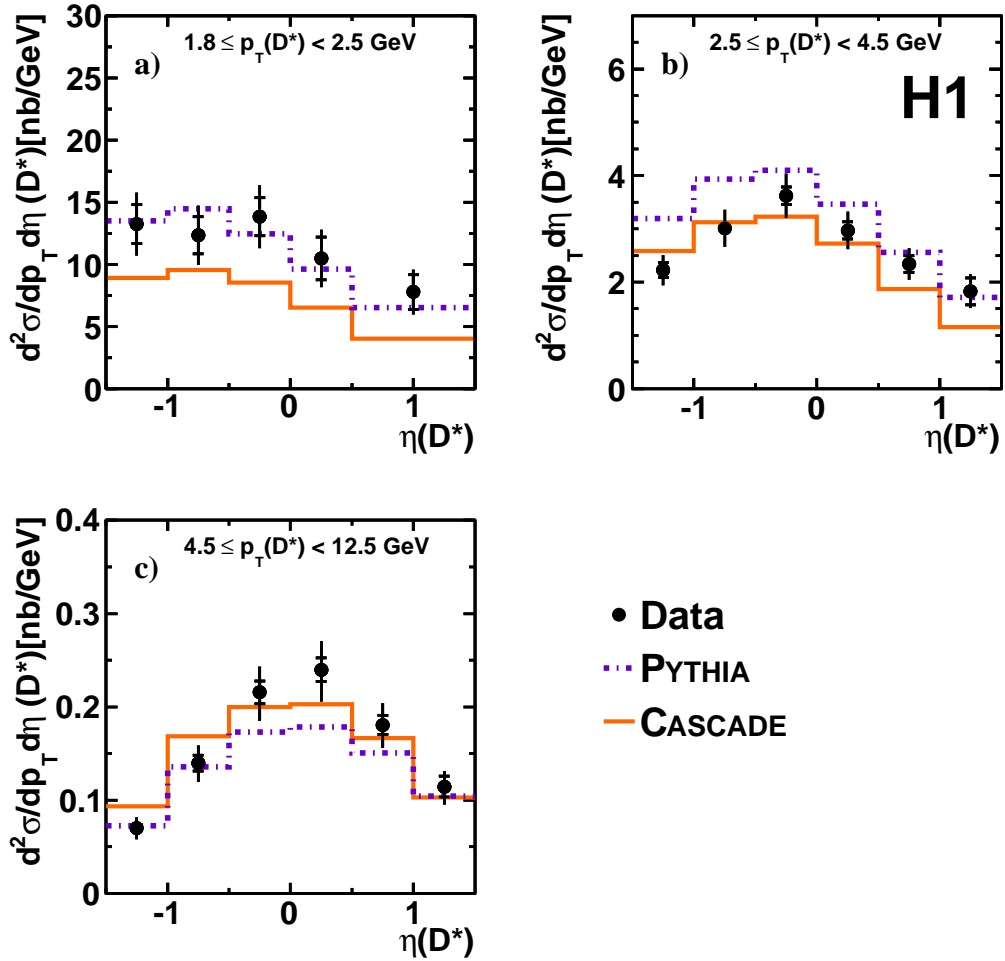}}
\put(2.,12.){\bf a)}
\put(9.,12.){\bf b)}
\put(2.,5.3){\bf c)}
\end{picture}
     \caption{Double differential \dstar\ cross section as a function of \etads\ for 
     three bins of \ptds\
    compared to \PYTHIA\ and \CASCADE\ predictions.}
    \label{ddiff}
\end{figure}

\begin{figure}[hp]
     \centering
\unitlength1.0cm
\begin{picture}(14,14)
\put(0,0){\includegraphics[width=14cm]{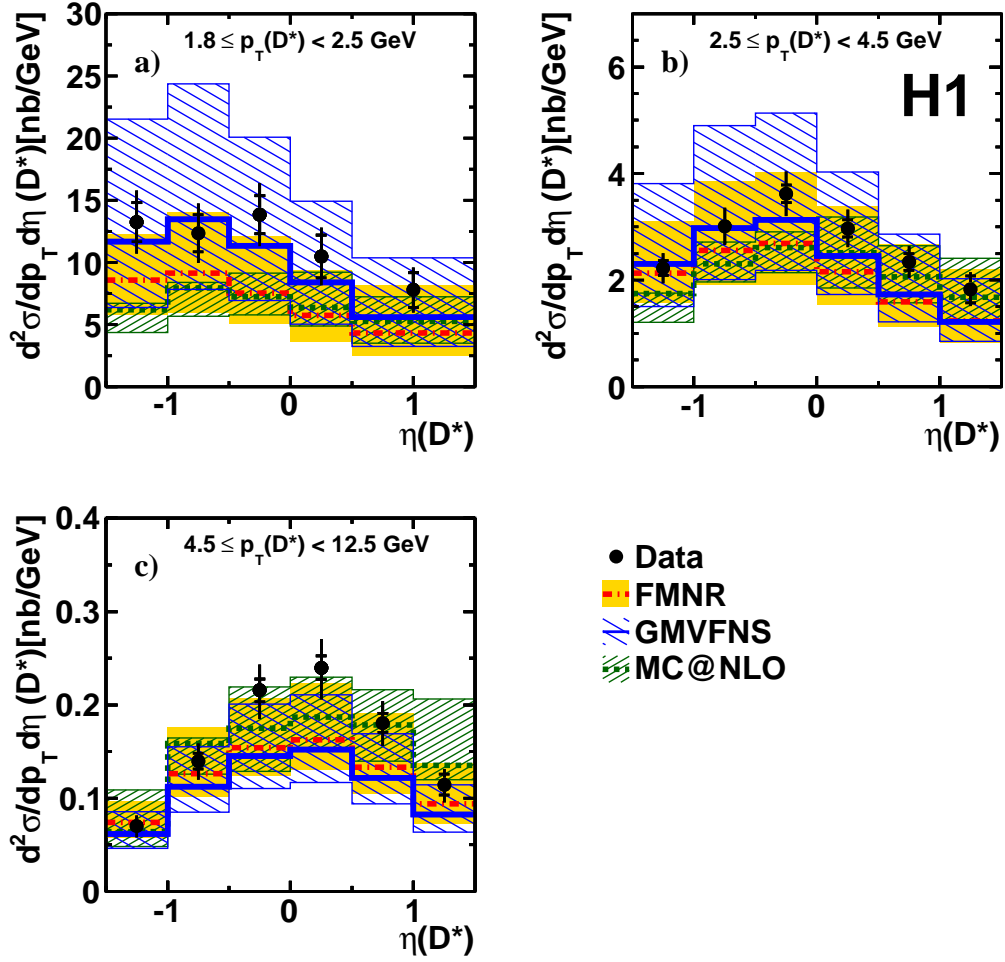}}
\put(2.,12){\bf a)}
\put(9.,12){\bf b)}
\put(2.,5.3){\bf c)}
\end{picture}
     \caption{Double differential \dstar\ cross section as a function of \etads\ for 
     three bins of \ptds\
    compared to the next-to-leading order predictions of 
    FMNR, GMVFNS and MC@NLO.}
    \label{ddiff_NLO}
\end{figure}

In order to investigate the correlation between pseudorapidity and  
transverse momentum, a double differential measurement in \ptds\ and \etads\ 
is performed (table~\ref{tab:csdd}). The cross sections of the 
leading order MCs \PYTHIA\ and 
\CASCADE\ in the three \ptds\ regions shown in figure~\ref{ddiff} reflect the 
different \ptds\ dependences seen in figure~\ref{singlediff}.
Both models are in broad agreement with the data.
The comparison of the NLO calculations with the data in 
figure~\ref{ddiff_NLO} leads to similar conclusions as for the LO MC
programs. 

\section{\boldmath Results for \dstar\ Tagged Dijet Production}

The integrated \dstarDj\  
cross section in the visible range 
given in table~\ref{tab:kinrange} is measured to be
\begin{equation}
  \label{eq:totXsecDstarjet}
  \sigma_{vis}(ep \to e \ \dstarjet +\mbox{\otherj} + X) = 
9.68 \pm 0.28~({\rm stat.}) \pm 0.51~({\rm unc. sys.}) 
\pm 0.64~({\rm norm.})~{\rm nb}.
\end{equation}
The corresponding predictions from \PYTHIA, \CASCADE\ and MC@NLO amount 
to $8.9~{\rm nb}$, $8.1~{\rm nb}$ and $7.1\,^{+2.5}_{-1.8}~{\rm nb}$, 
respectively. 
In the common range of transverse momentum, $\ptds>2.1\ {\rm GeV}$, the 
ratio of the \dstarDj\ to the inclusive \dstar\ cross section is
$0.304 \pm 0.013 \pm 0.031$, compared to $0.271$ and $0.311$
for \PYTHIA\ and \CASCADE, respectively.
MC@NLO predicts a ratio of $0.309\,^{+0.019}_{-0.040}$.

\begin{figure}[ht]
\unitlength1.0cm
\begin{picture}(16,18)
\put(0,6.5){\includegraphics[width=8cm]{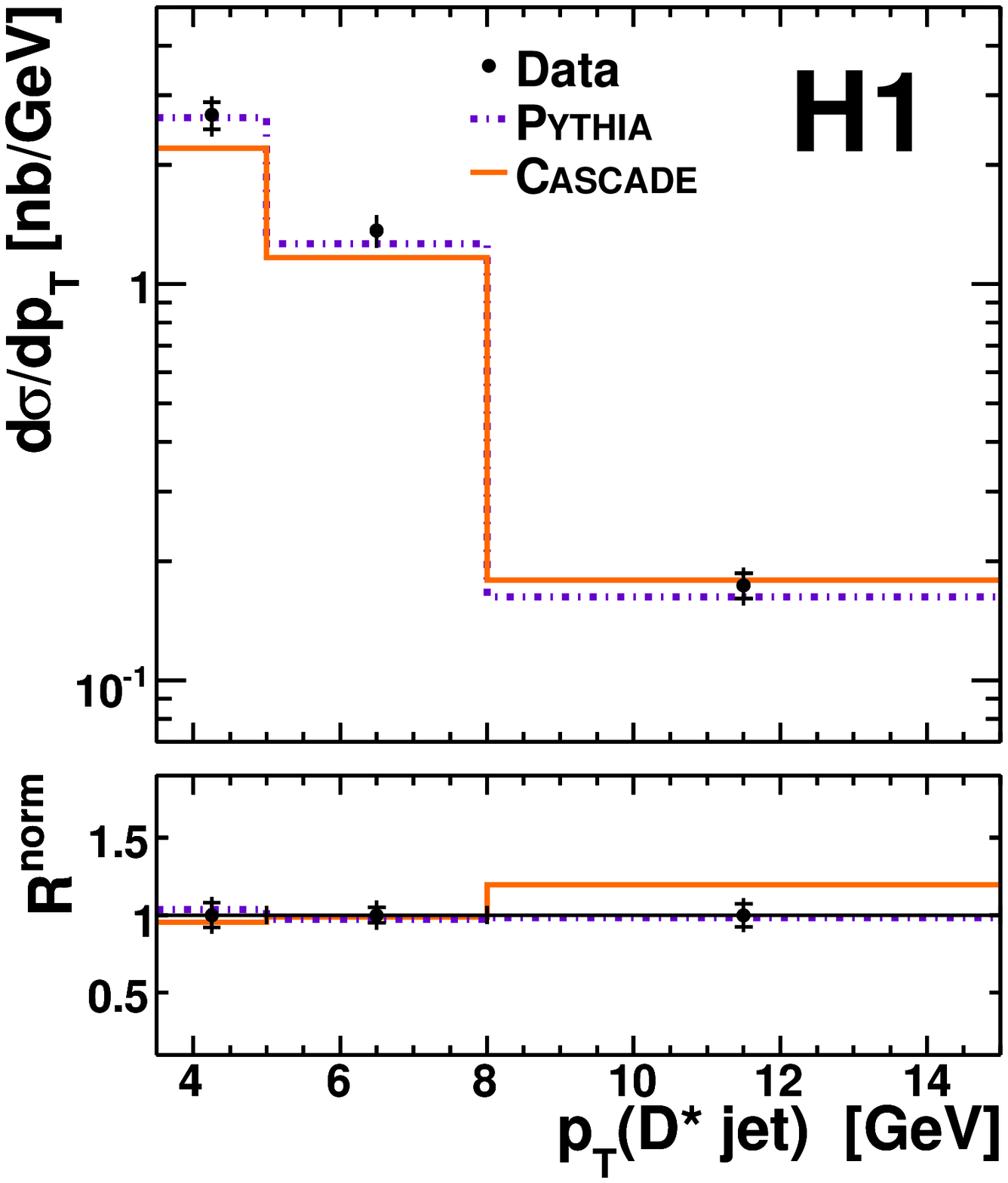}}
\put(8,6.5){\includegraphics[width=8cm]{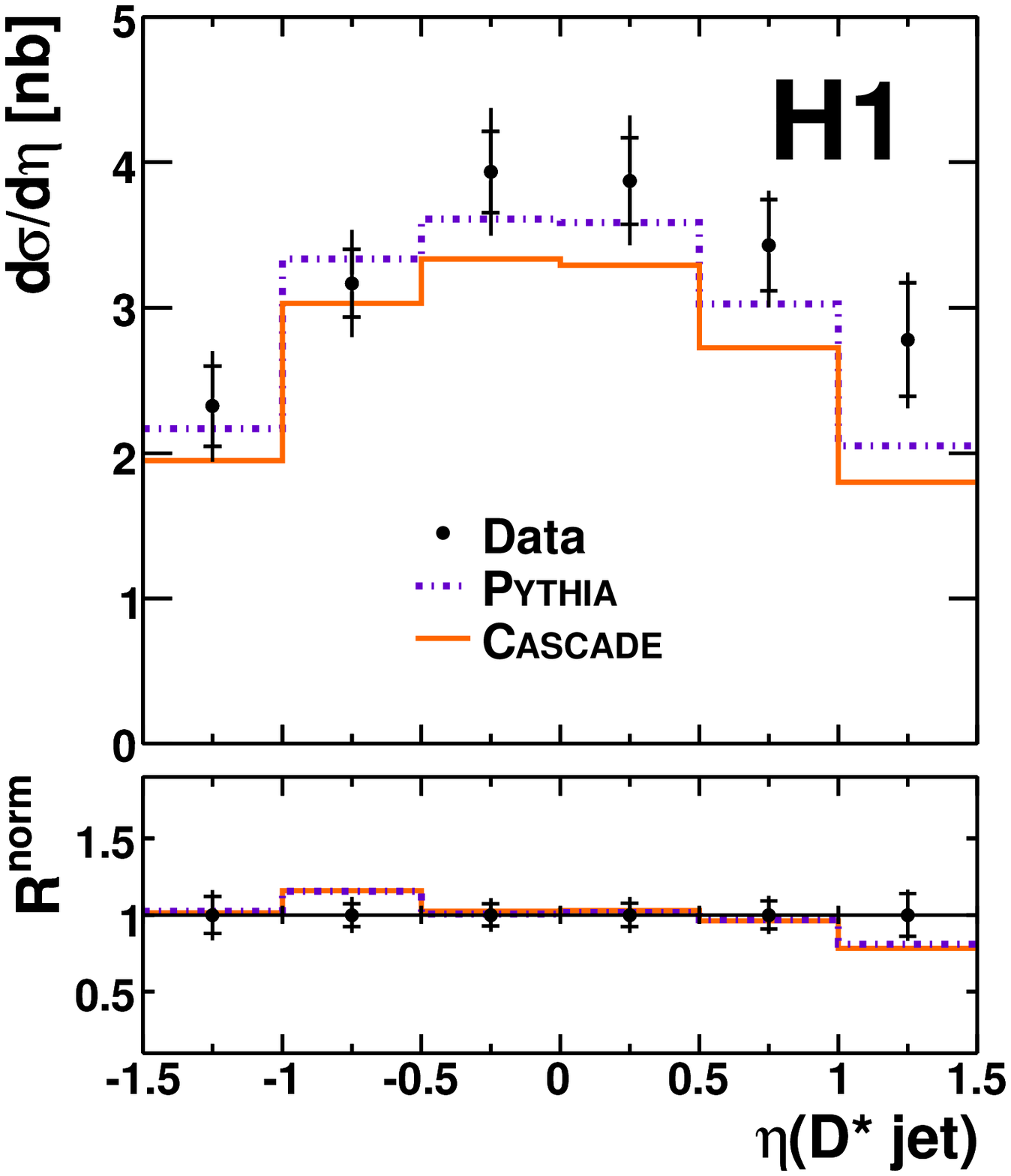}}
\put(0,-3){\includegraphics[width=8cm]{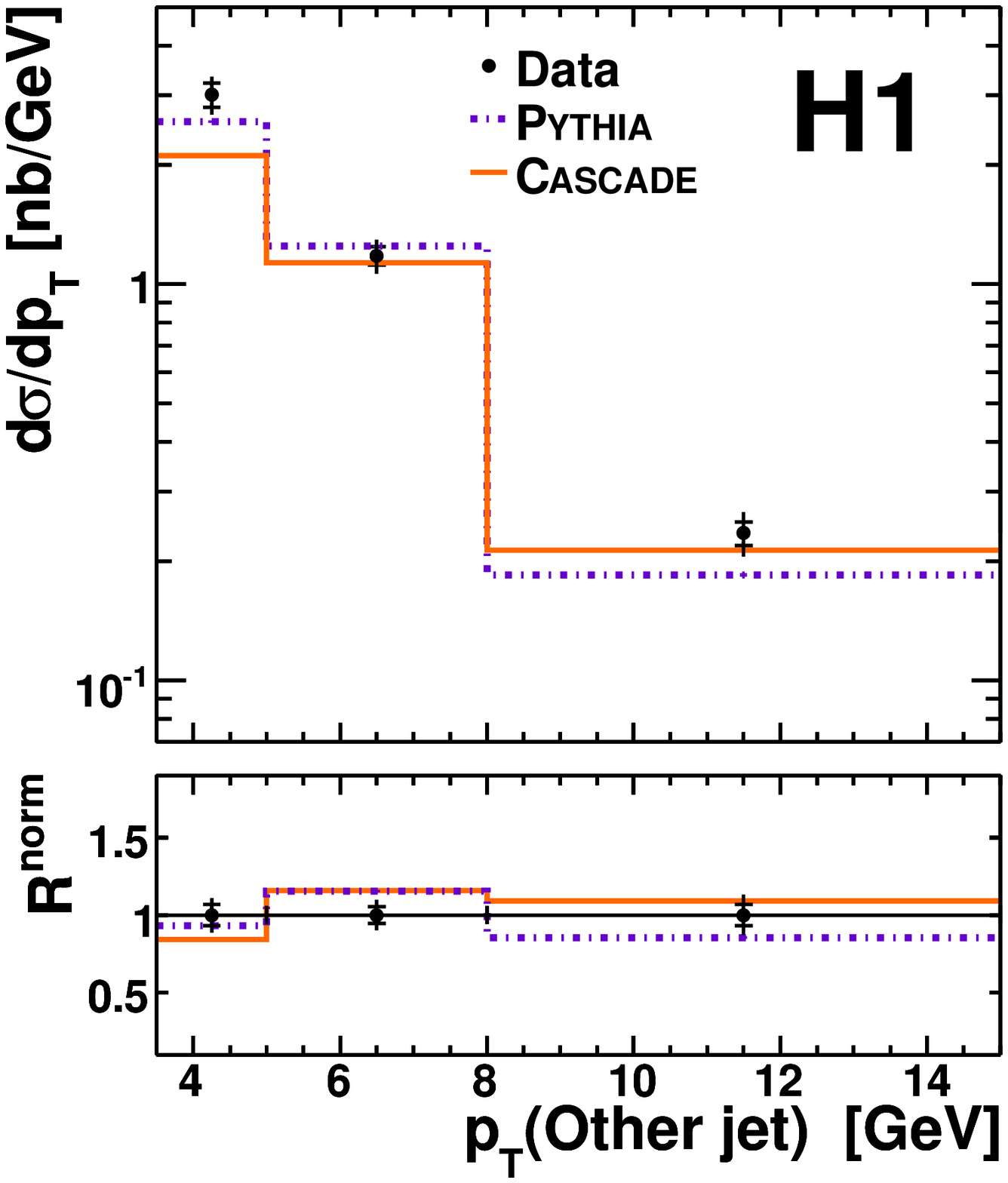}}
\put(8,-3){\includegraphics[width=8cm]{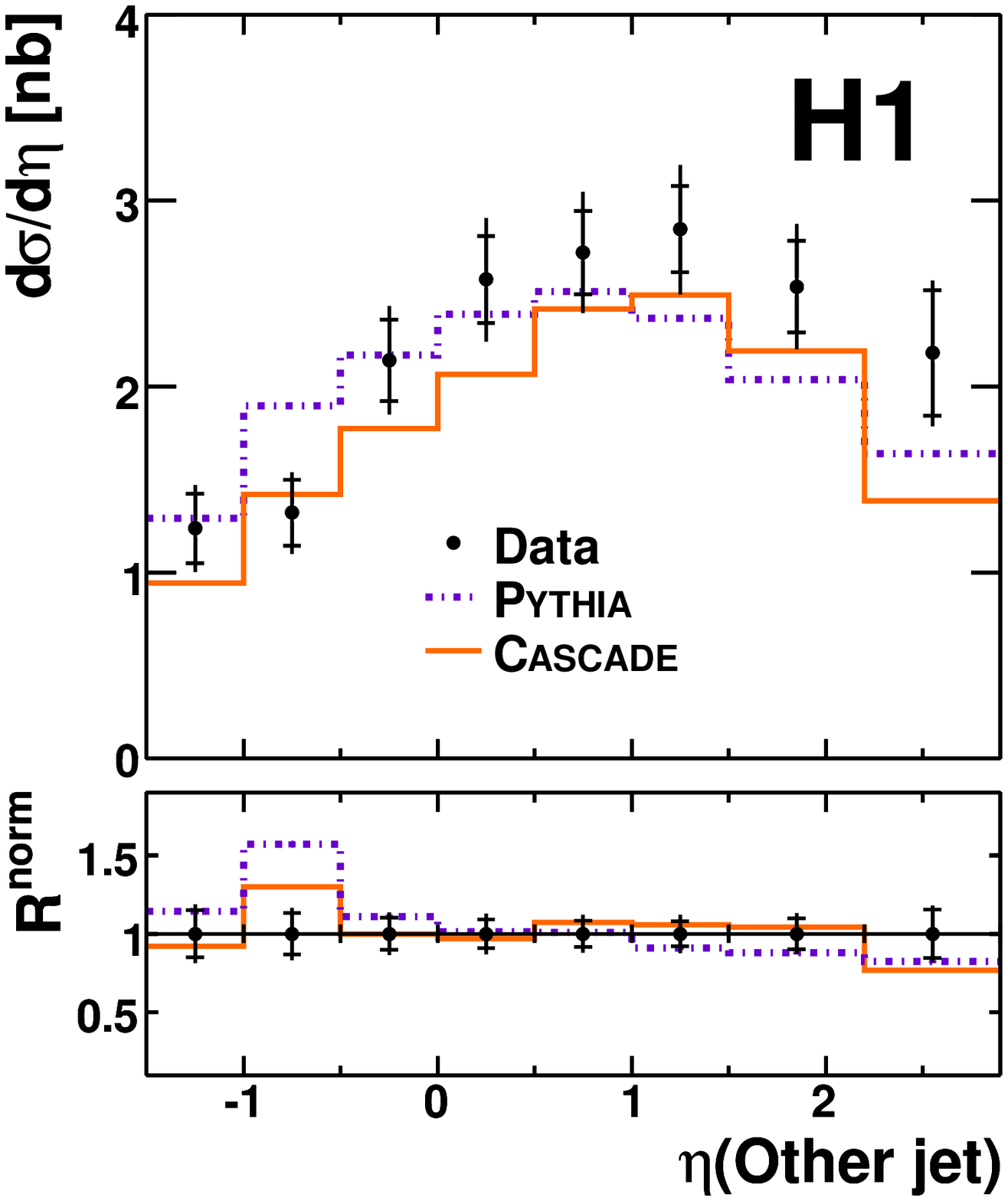}}
\put(2.5,17.5){\bf a)}
\put(9.8,17.5){\bf b)}
\put(2.5,8){\bf c)}
\put(9.8,8){\bf d)}
\end{picture}
   \caption{Single differential cross section for \dstarDj\ production 
    as a function of $\pt$ and $\eta$ of the \dstarjet\  and the \otherj\  
    compared to \PYTHIA\ and \CASCADE\ predictions.  The normalised 
    ratio \rnorm\ (see text) is also shown.}
    \label{fig:dstardijet}
\end{figure}

\begin{figure}[ht]
\unitlength1.0cm
\begin{picture}(16,18)
\put(0,6.5){\includegraphics[width=8cm]{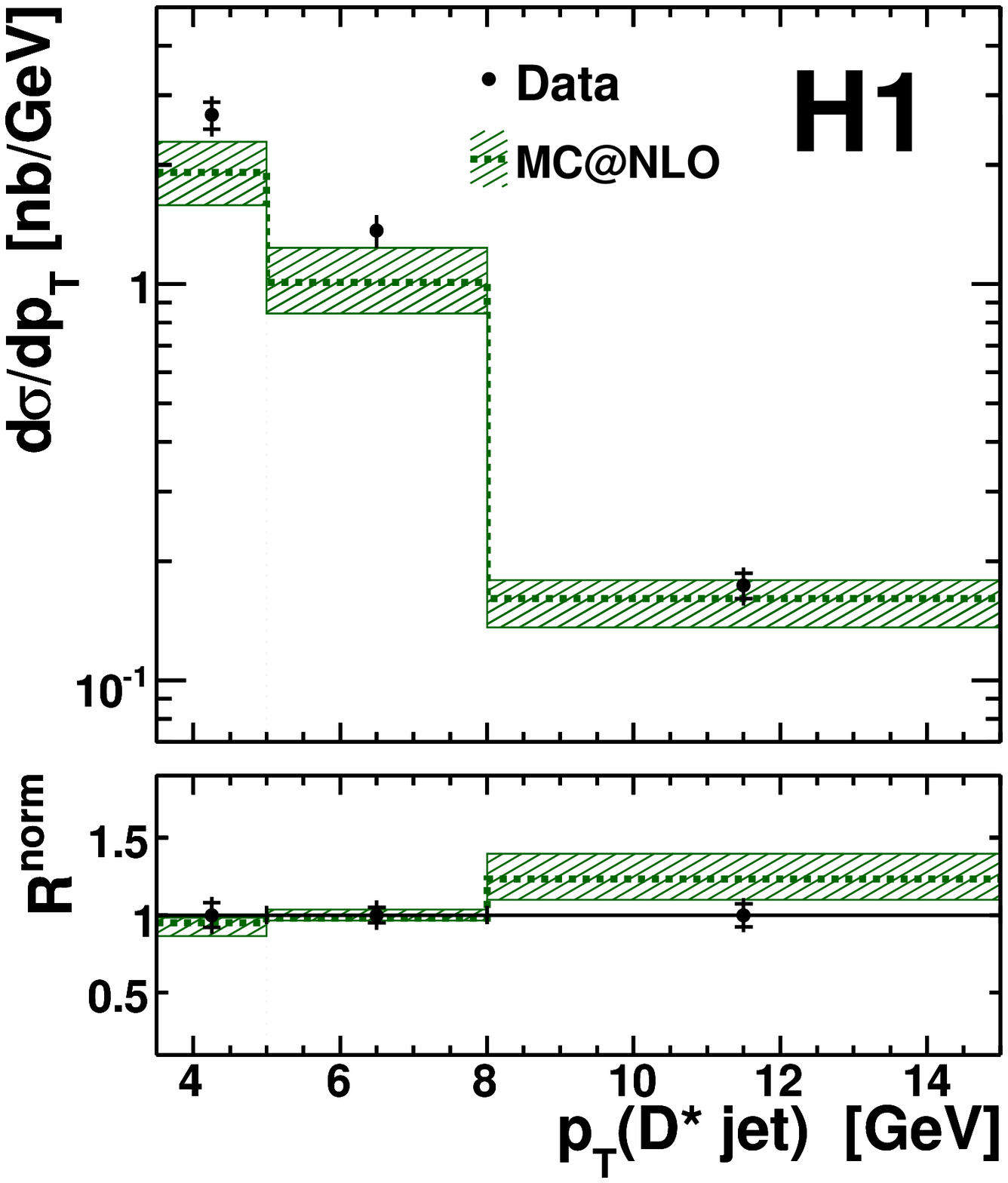}}
\put(8,6.5){\includegraphics[width=8cm]{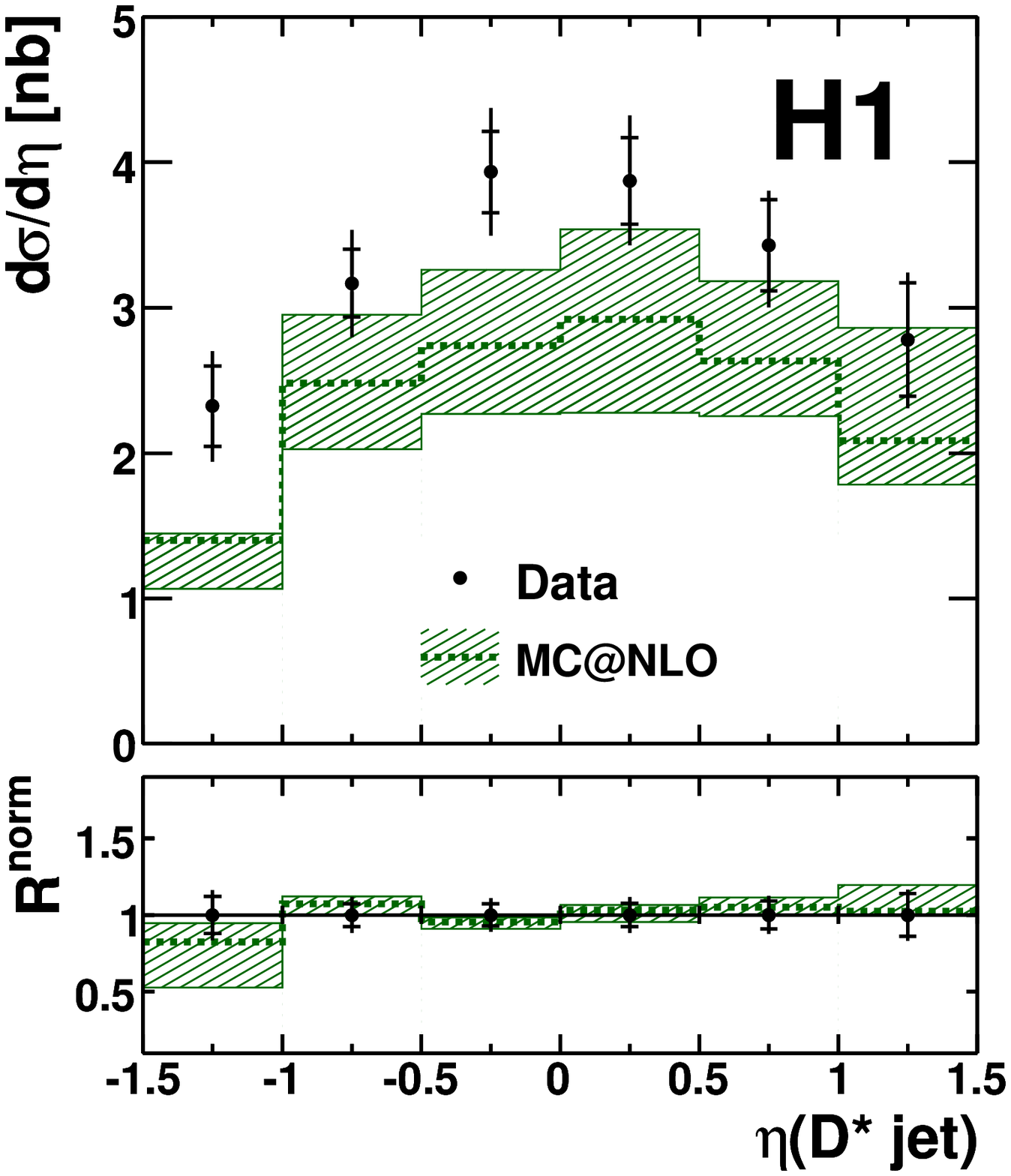}}
\put(0,-3){\includegraphics[width=8cm]{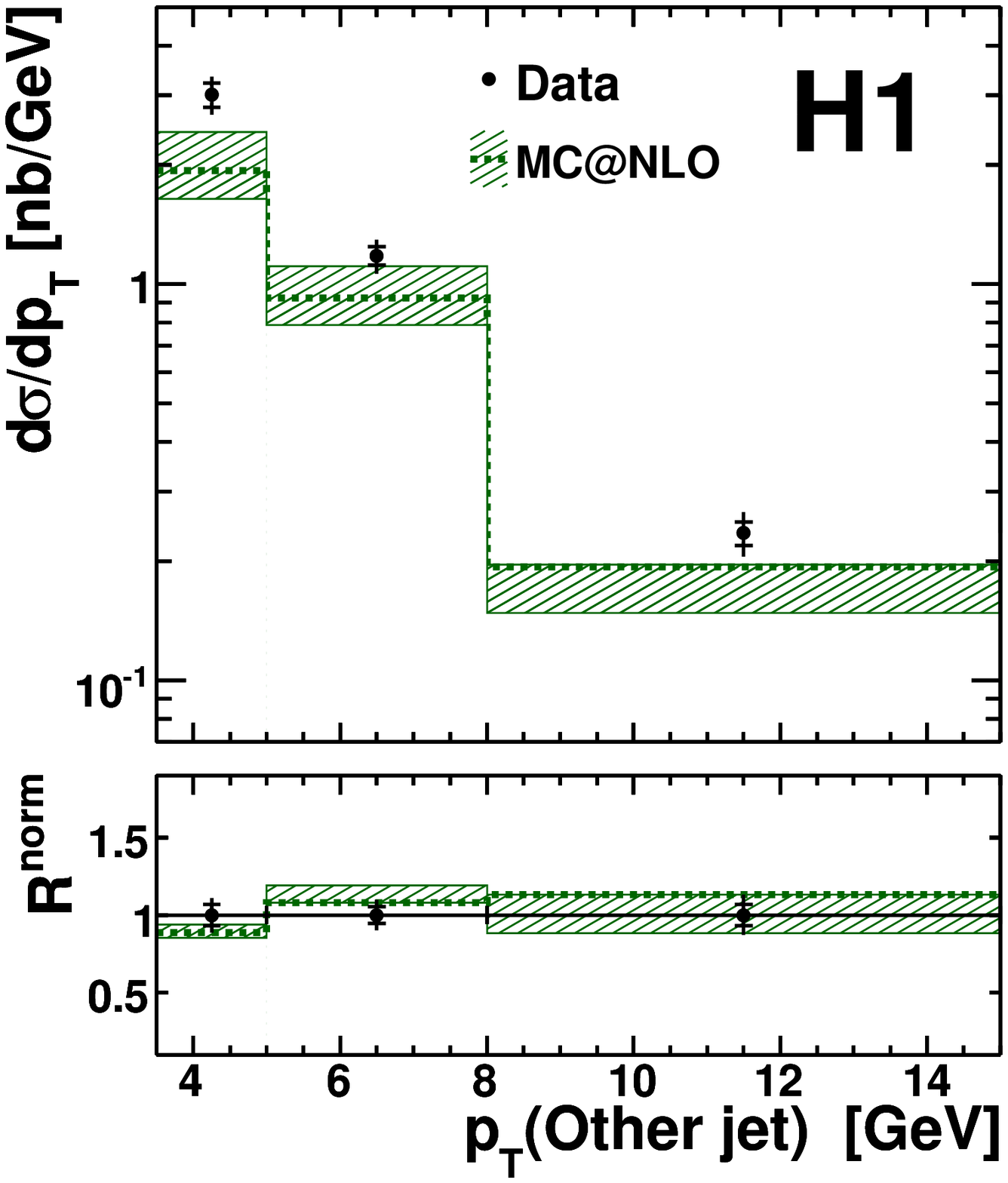}}
\put(8,-3){\includegraphics[width=8cm]{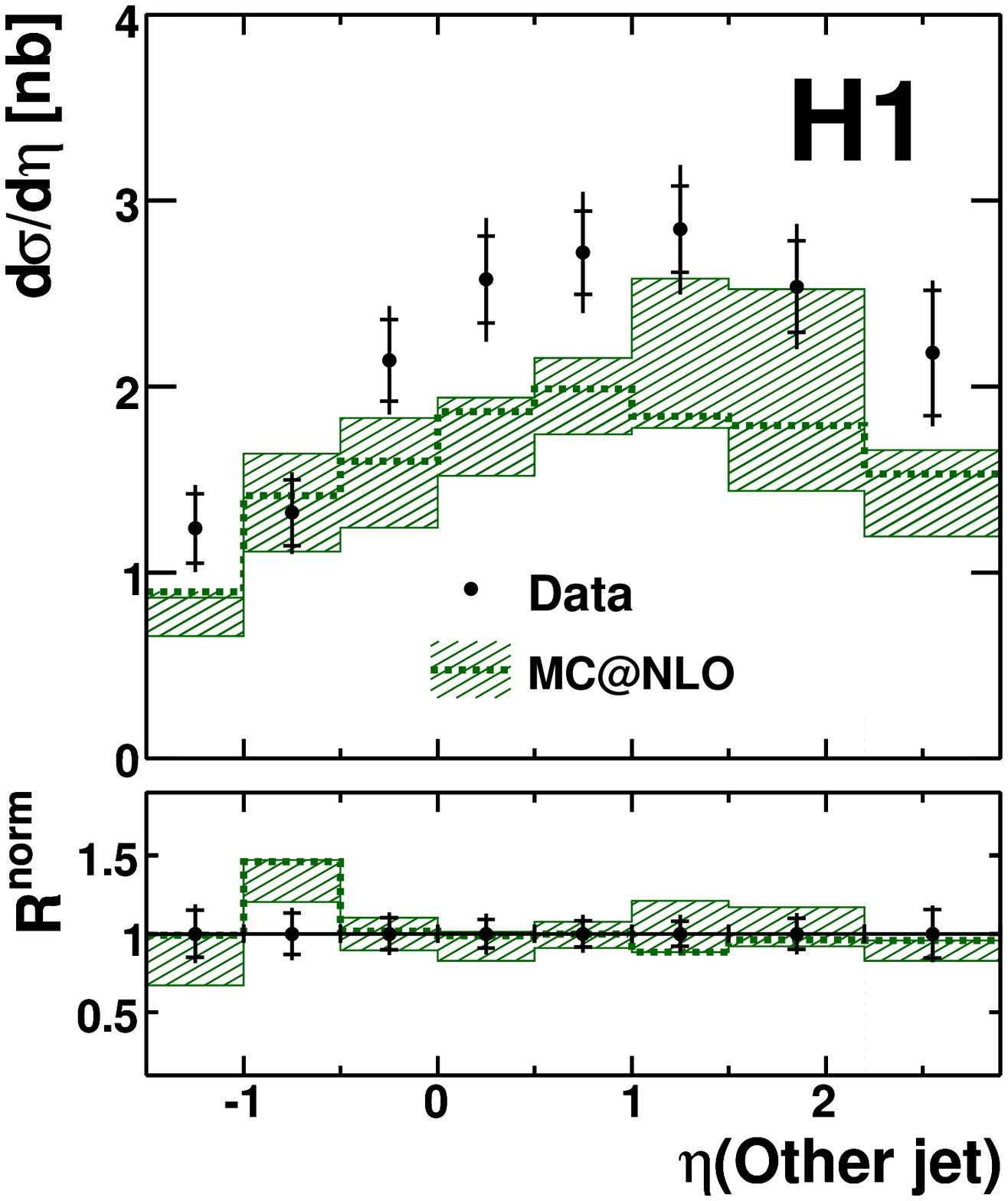}}
\put(2.5,17.5){\bf a)}
\put(9.8,17.5){\bf b)}
\put(2.5,8){\bf c)}
\put(9.8,8){\bf d)}
\end{picture}
   \caption{Single differential cross section for \dstarDj\ production 
    as a function of $\pt$ and $\eta$ of the \dstarjet\  and the \otherj\  
    compared to MC@NLO predictions. The normalised 
    ratio \rnorm\ (see text) is also shown. }
    \label{fig:dstardijet-mcatnlo}
\end{figure}

The bin averaged differential cross section for the \dstarDj\ production 
as a function 
of the transverse momentum $\pt$ and the pseudorapidity $\eta$ of both the 
\dstarjet\ and the \otherj\ are 
listed in table~\ref{tab:diffXsecdijet}
and shown in figures~\ref{fig:dstardijet} and~\ref{fig:dstardijet-mcatnlo}. 
On average, the \otherj\ is more forward than the \dstarjet\ not only due
to the larger measurement range in $\eta$, but also within the common region
of $-1.5 < \eta< 1.5$. This behaviour 
is consistent with the expectation that the \otherj\ originates not always from
a charm quark.  This observation confirms the 
result of the previous H1 analysis of \dstarDj\ photoproduction~\cite{h1dstar06}
with improved precision.
In figure~\ref{fig:dstardijet} the measurements are compared to the \PYTHIA\ 
and the \CASCADE\ predictions. The shapes of the distributions are described 
well by both models. 
In figure~\ref{fig:dstardijet-mcatnlo} the measurements are compared to the 
predictions of MC@NLO. At low transverse momenta of both the \dstarjet\ and the 
\otherj, the predictions lie significantly below the measurement. This 
results in a smaller total visible cross section which is also observed in 
the $\eta$ distribution. 
The uncertainty band of the MC@NLO prediction includes both variation of the 
charm mass and variations of the factorisation and renormalisation scales as 
described in section~\ref{mcatnlo-description}.

In order to investigate further the charm production dynamics, several 
variables related to the structure of the hadronic final state are studied.
The correlation between the jets in the longitudinal and transverse 
directions is experimentally assessed by the difference in pseudorapidity
$\Delta \eta=\eta(\otherj)-\eta(\dstarjet)$ and
in the azimuthal angle \dphidsj\ between the \dstarjet\ and 
the \otherj. The amount of QCD radiation in addition to the
the two leading jets is investigated using the mass variable
$M_X = \sqrt{(P + q  -  (j_1 + j_2 ))^2}$  with $P$, $q$, $j_1$ and $j_2$ being 
the four-vectors of the initial proton, the exchanged photon, the \dstarjet\
and the \otherj, respectively. In direct photon processes without 
radiation, $M_X$ is expected to be close to the
proton mass, whereas resolved processes as well as additional QCD radiation 
will increase $M_X$.
The fraction $x_\gamma$ of the longitudinal photon momentum 
entering the hard scattering process can be used to distinguish
direct and resolved processes:
in collinear factorisation at LO a resolved photon process is characterised by
$x_\gamma < 1$, while a direct process has $x_\gamma = 1$.
In the \dstarDj\ sample, $x_\gamma$ is approximated by
\begin{equation}
  \xgjj  = \frac{\sum_{jets} (E -p_z)_i}{\sum_{HFS} (E-p_z)_j}.
\end{equation}
The sum in the numerator runs over the particles in the two selected jets,
whereas the sum in the denominator contains all reconstructed
particles of the hadronic final state.
 
\begin{figure}[ht]
\unitlength1.0cm
\begin{picture}(16,18)
\put(0,6.5){\includegraphics[width=8cm]{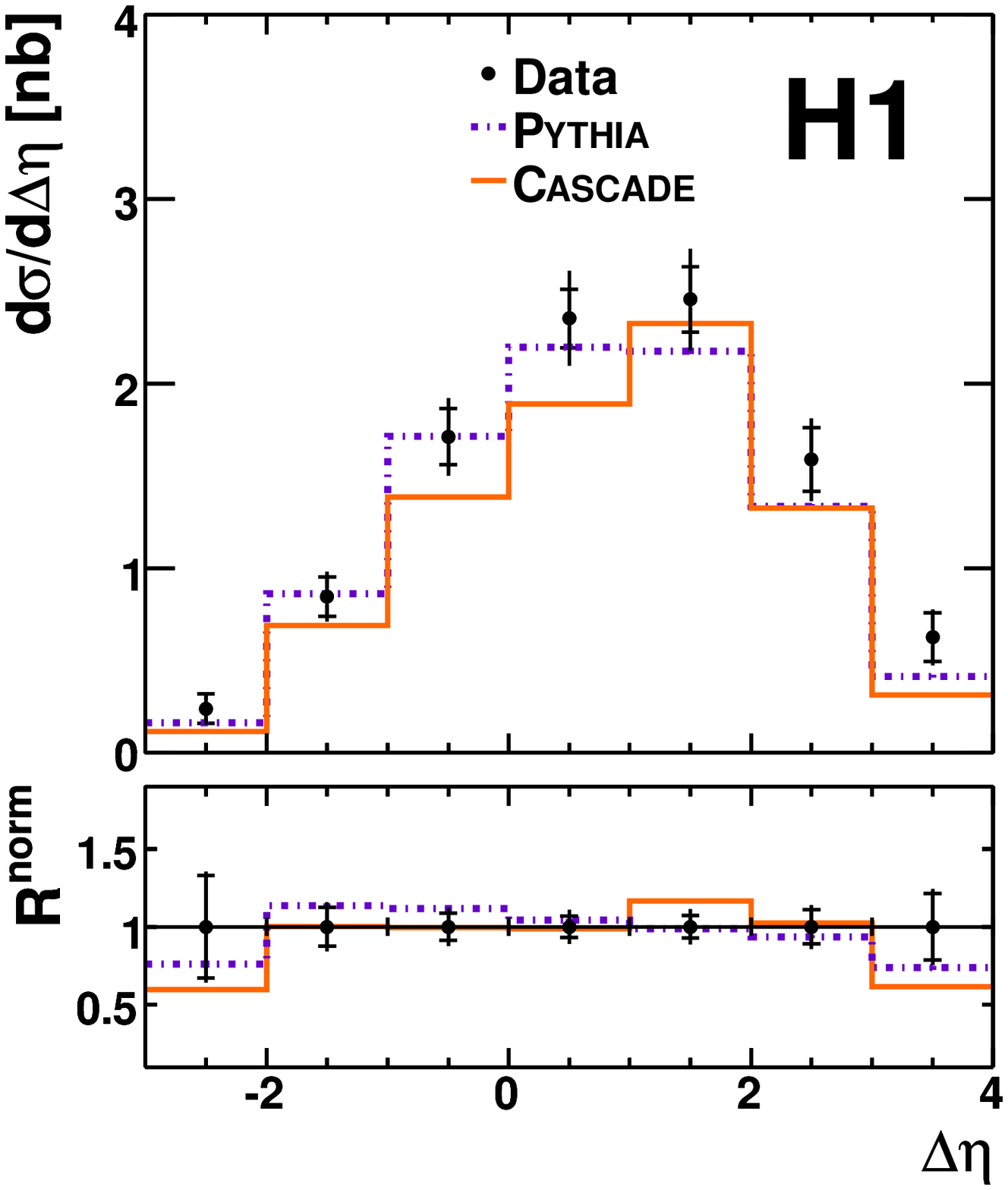}}
\put(8,6.5){\includegraphics[width=8cm]{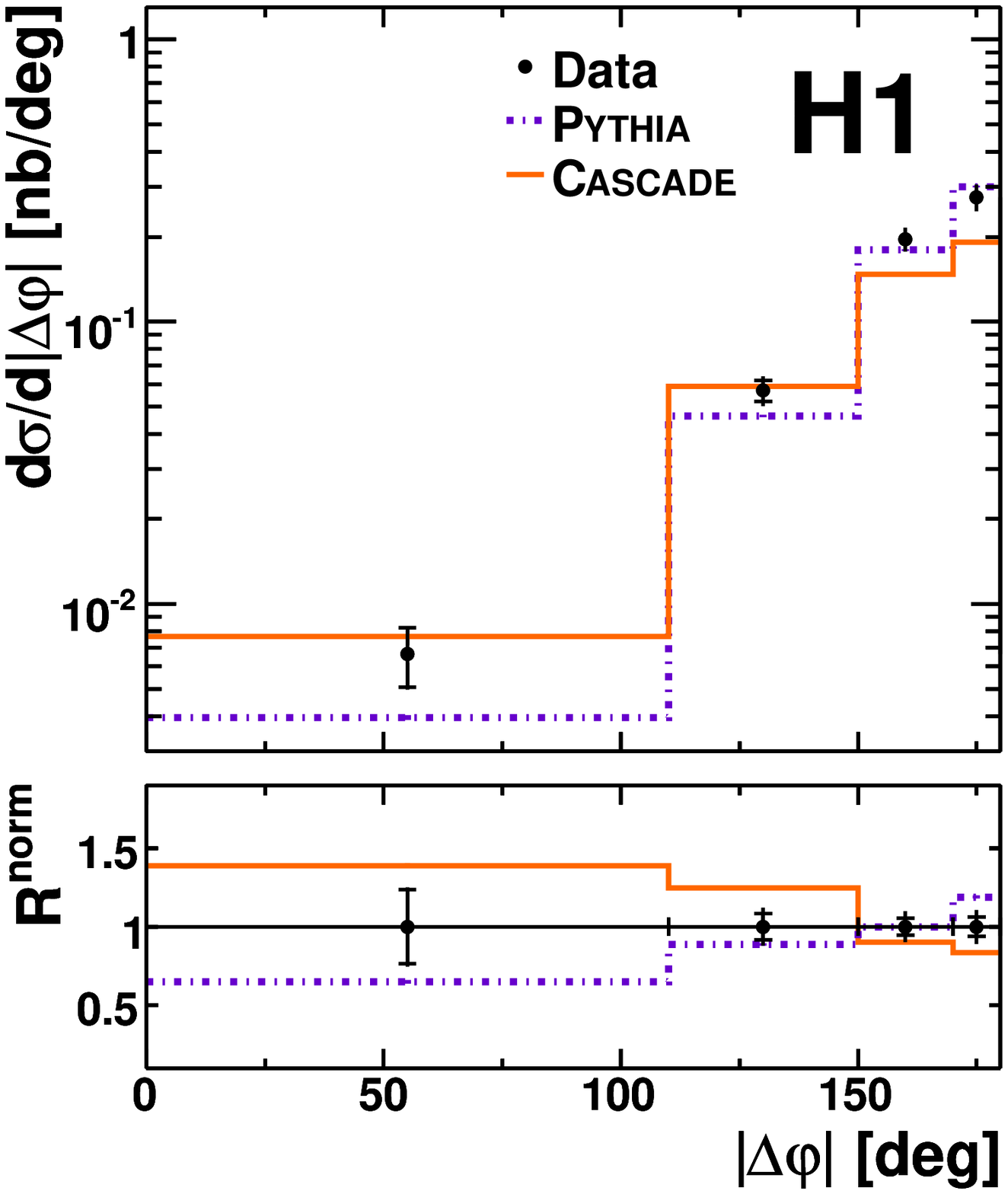}}
\put(0,-3){\includegraphics[width=8cm]{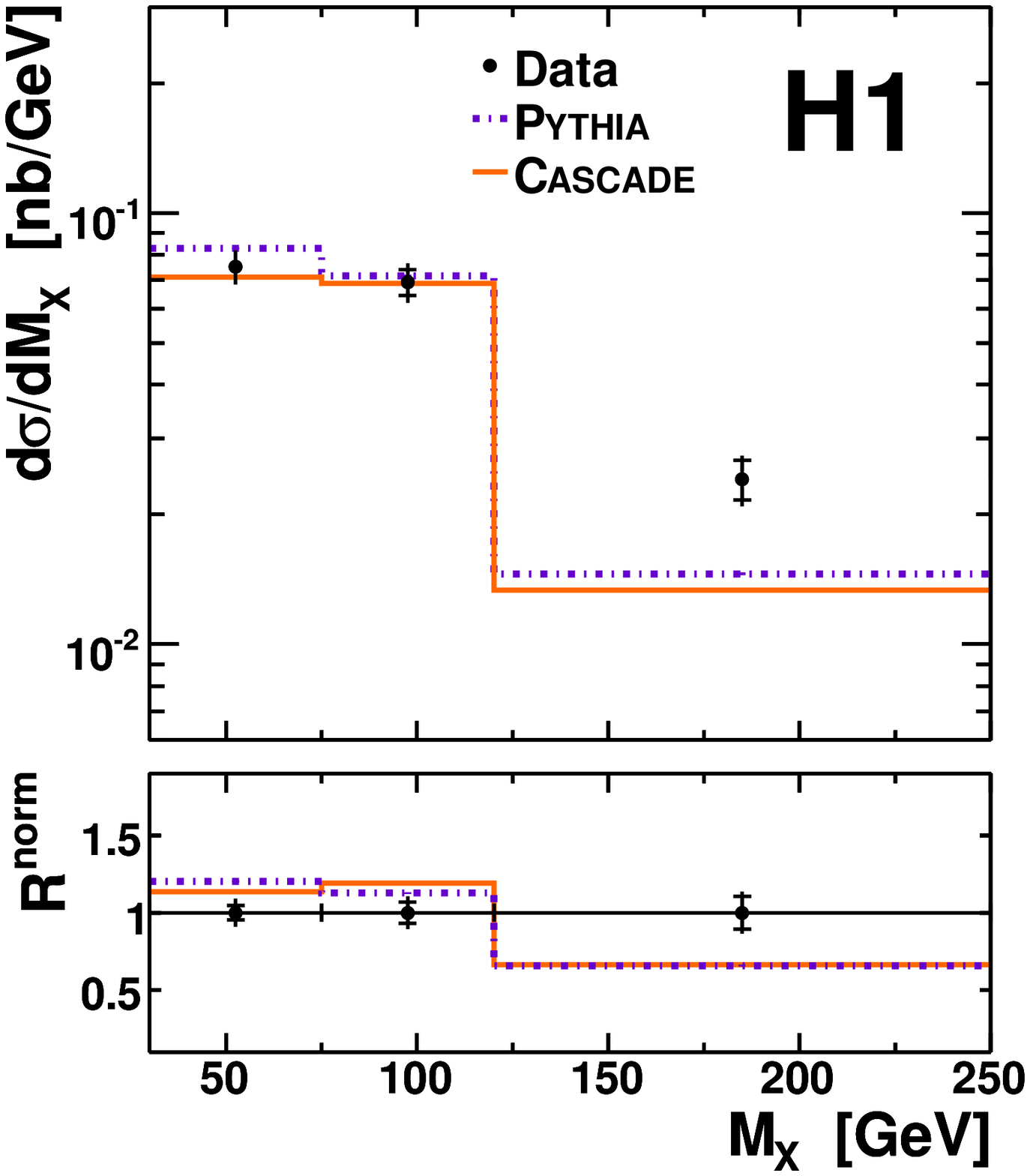}}
\put(8,-3){\includegraphics[width=8cm]{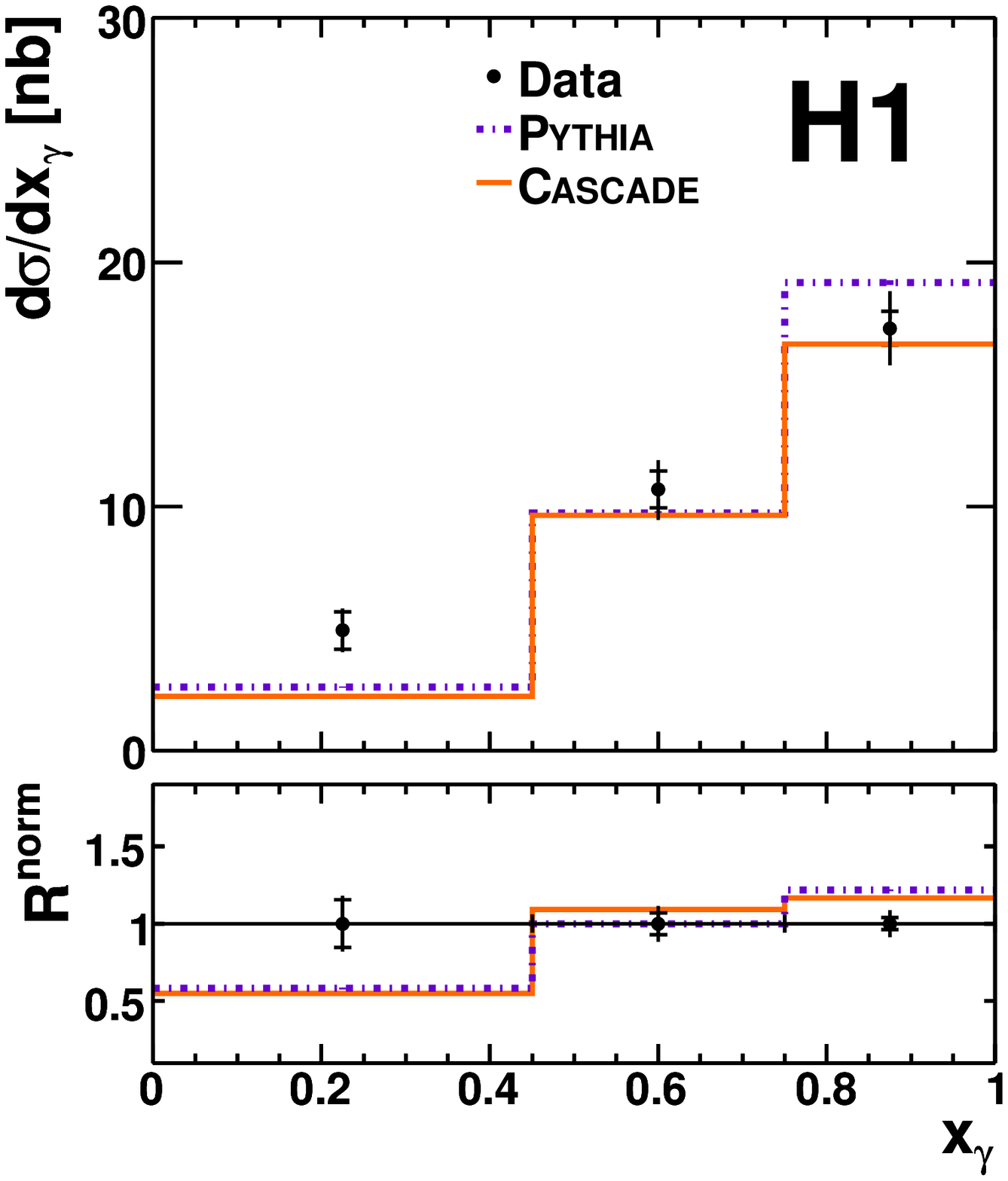}}
\put(1.8,17.5){\bf a)}
\put(9.8,17.5){\bf b)}
\put(1.8,8){\bf c)}
\put(9.8,8){\bf d)}
\end{picture}
   \caption{Single differential cross section for \dstarDj\ production 
   as a function 
   of the difference in pseudorapidity $\Delta \eta$ and in azimuthal 
   angle $\Delta \varphi$ between the \otherj\ and the \dstarjet, the mass $M_X$ 
   and $\xgjj$   
     compared to \PYTHIA\ and \CASCADE\ predictions. The normalised ratio \rnorm\
    (see text) is also shown.}
    \label{fig:dstardijet-corr}
\end{figure}

\begin{figure}[ht]
\unitlength1.0cm
\begin{picture}(16,18)
\put(0,6.5){\includegraphics[width=8cm]{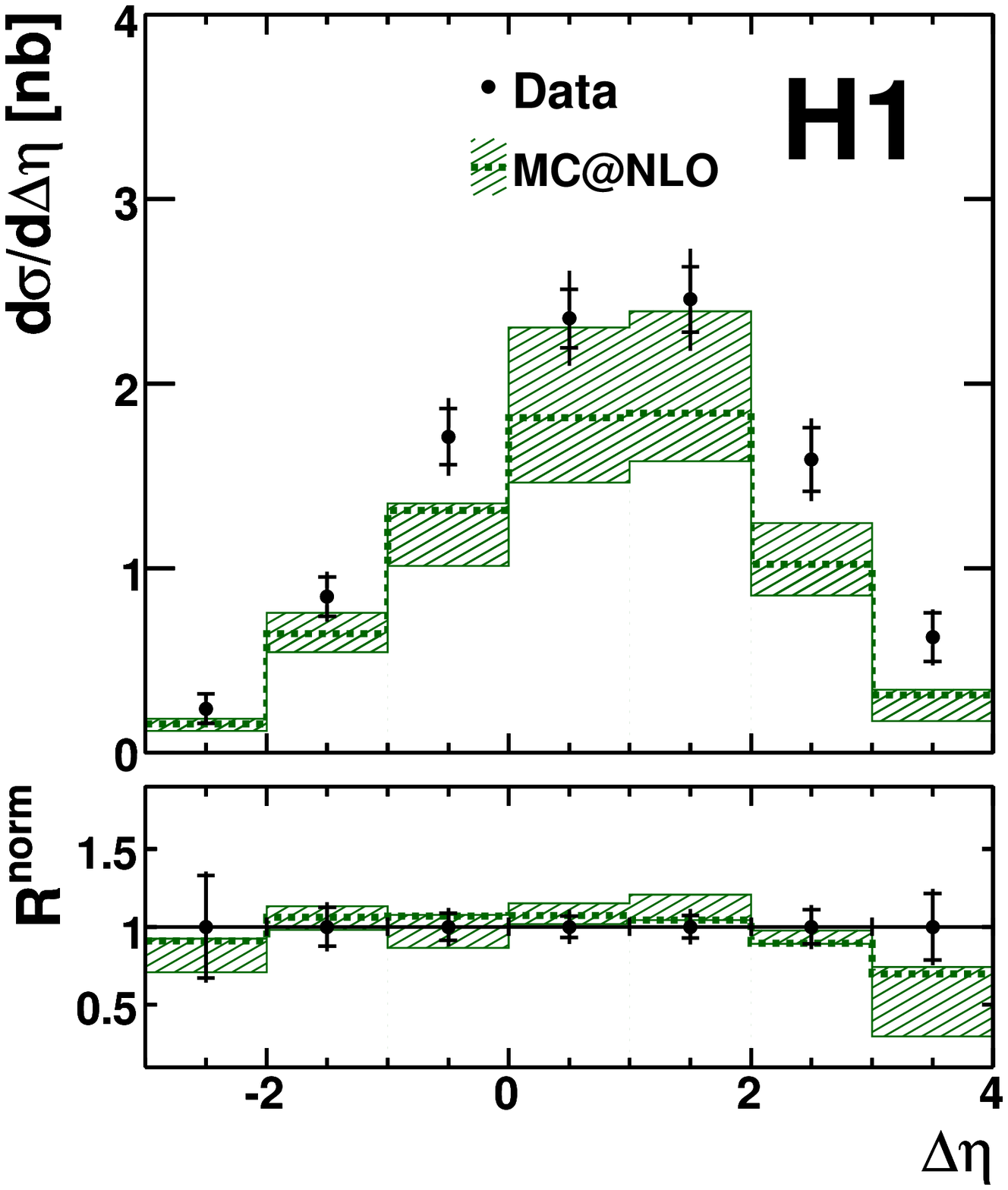}}
\put(8,6.5){\includegraphics[width=8cm]{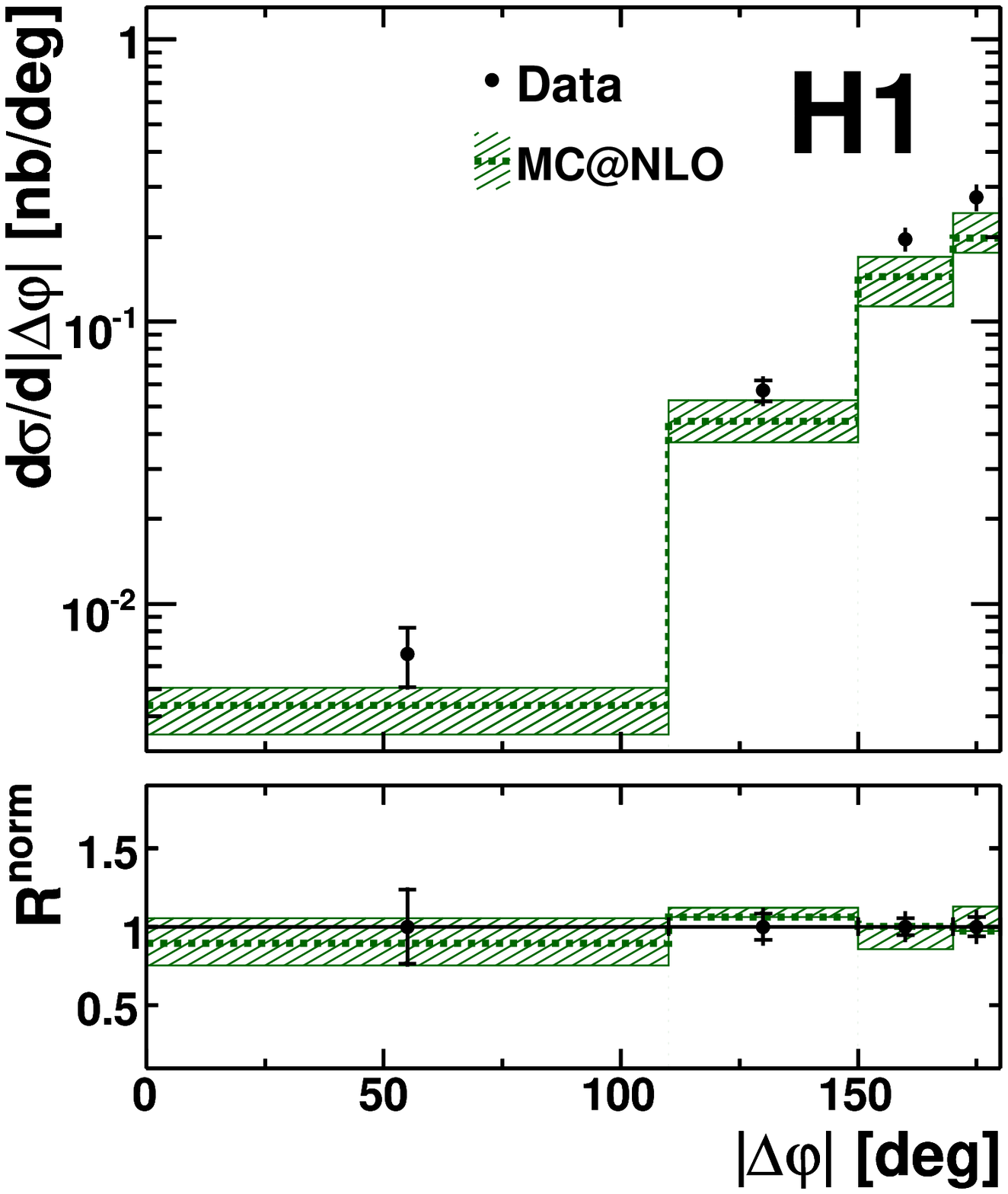}}
\put(0,-3){\includegraphics[width=8cm]{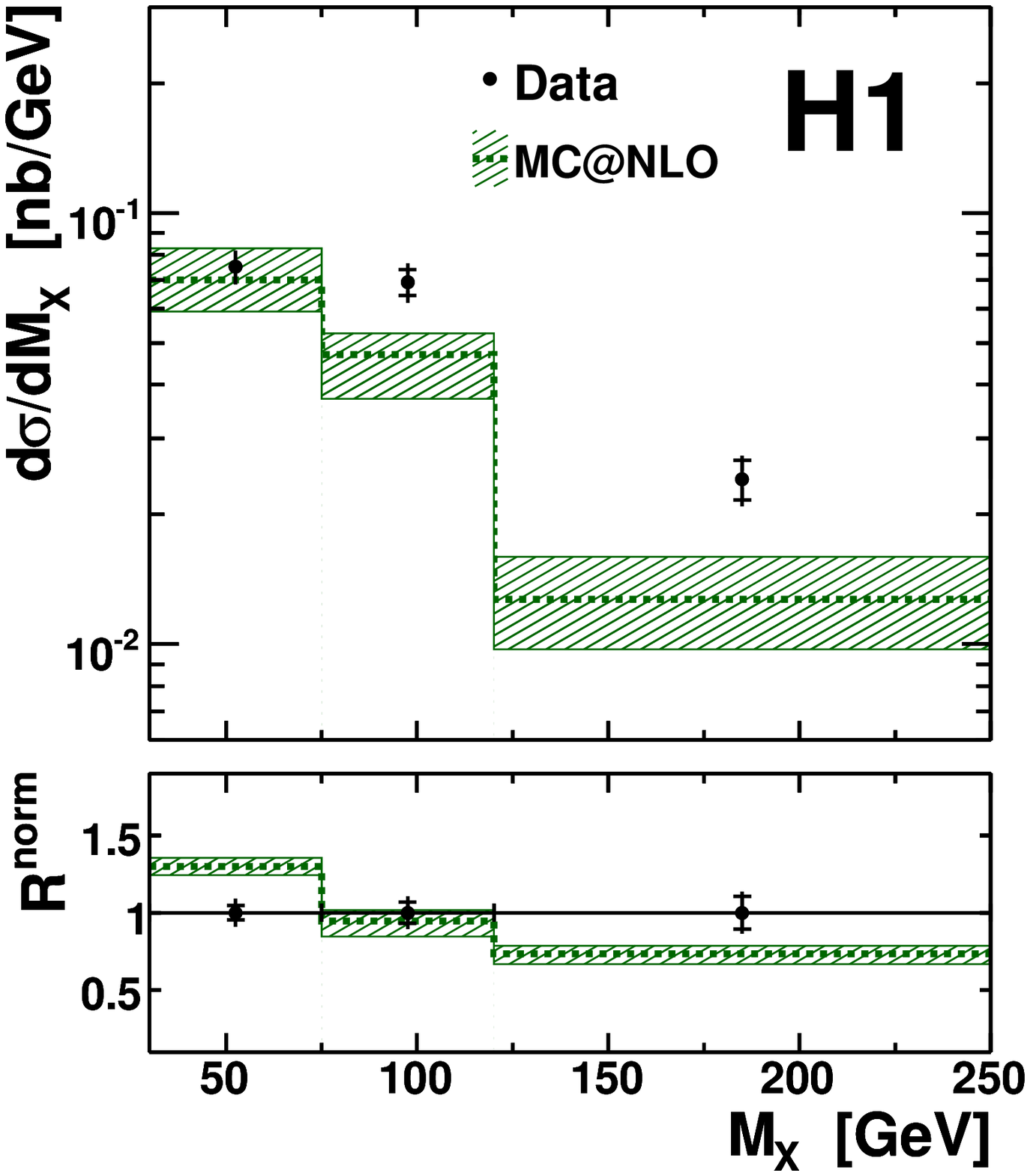}}
\put(8,-3){\includegraphics[width=8cm]{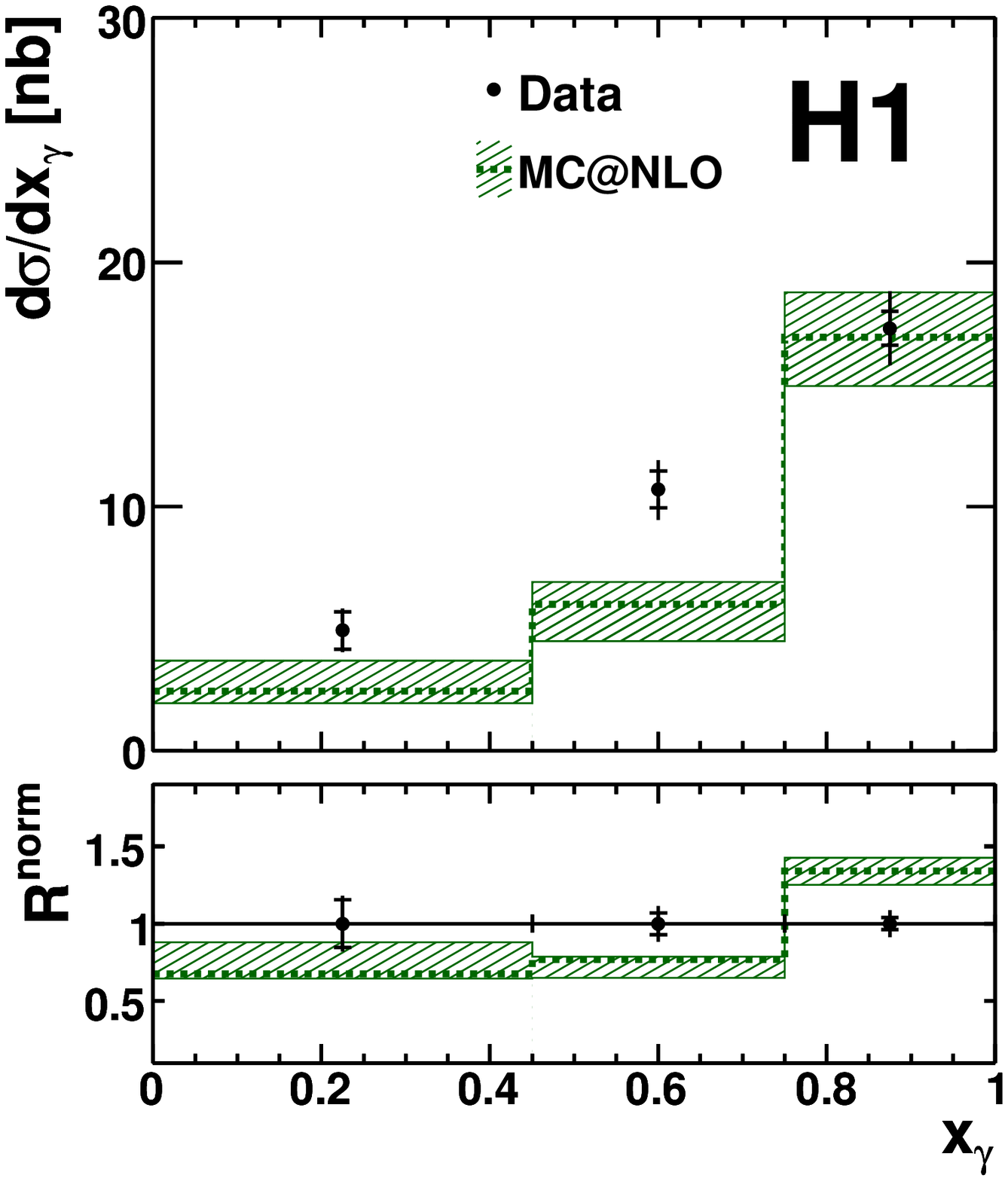}}
\put(1.8,17.5){\bf a)}
\put(9.8,17.5){\bf b)}
\put(1.8,8){\bf c)}
\put(9.8,8){\bf d)}
\end{picture}
   \caption{Single differential cross section for \dstarDj\ production as a function 
   of the difference in pseudorapidity $\Delta \eta$ and in azimuthal 
   angle $\Delta \varphi$ between the \otherj\ and the \dstarjet, the mass $M_X$ 
   and $\xgjj$   
    compared to MC@NLO predictions. The normalised ratio \rnorm\
    (see text) is also shown.}
    \label{fig:dstardijet-corr-mcatnlo}
\end{figure}

In table~\ref{tab:diffXsecdijetcorr} and
figures~\ref{fig:dstardijet-corr} and~\ref{fig:dstardijet-corr-mcatnlo} the 
bin averaged differential cross sections for the \dstarDj\ production as a 
function of the difference in pseudorapidity 
$\Delta \eta$ and in azimuthal 
angle \dphidsj\ between the \otherj\ and the \dstarjet, the mass $M_X$ 
and $\xgjj$ are presented.
The cross section as a function of $\Delta \eta$ is not symmetric because 
the \otherj\ is on average more forward than the \dstarjet. 
The shape in $\Delta \eta$  is 
reasonably well described by all QCD calculations.
The cross section as a function of \dphidsj\ shows a significant 
contribution away from the back-to-back configuration at  
$|\Delta \varphi| \simeq 180 \grad$. Such a configuration can be described 
by models which include significant contributions 
from higher order QCD radiation or a transverse momentum of the gluon in the
initial state. 
Whereas \PYTHIA\ predicts a too small relative contribution of these
configurations, \CASCADE\ overestimates them. 
The prediction from MC@NLO, shown in 
figure~\ref{fig:dstardijet-corr-mcatnlo}b), agrees 
well in shape with the measurement. 

The cross section as a function of the invariant mass $M_X$ is reasonably 
well described by the predictions of \CASCADE\ and \PYTHIA\ in the region 
of $M_X < 120$~GeV, whereas the measured cross section is larger than the 
predictions for the highest $M_X$ bin. The large $M_X$ region is correlated 
with the region of small $\xgjj$, where also the predictions are below the 
measurement.
MC@NLO predicts a different shape for $M_X$ and 
is not able to describe the shape of the $\xgjj$ distribution.

\begin{figure}[ht]
\unitlength1.0cm
\begin{picture}(16,18)
\put(0,6.5){\includegraphics[width=8cm]{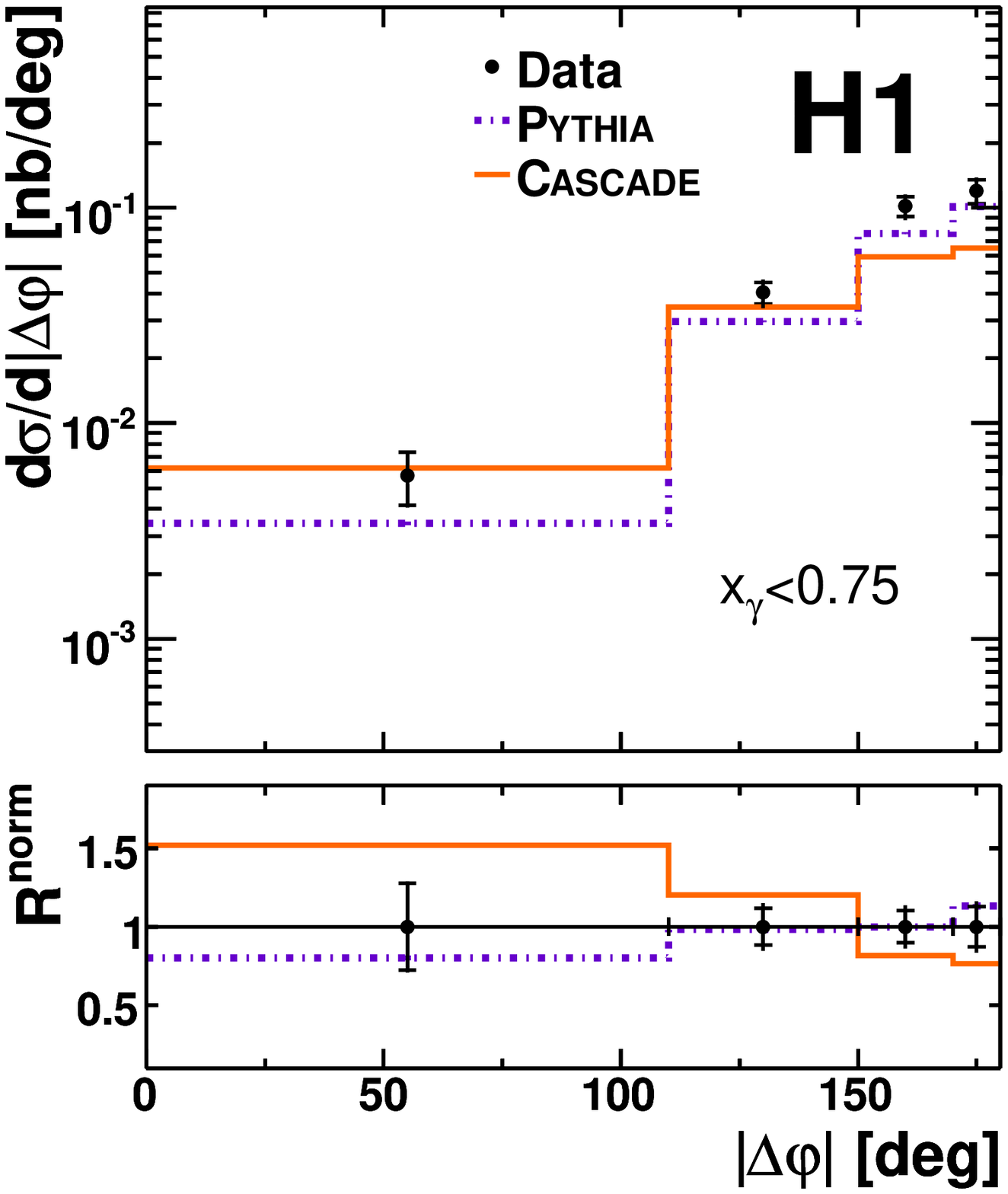}}
\put(8,6.5){\includegraphics[width=8cm]{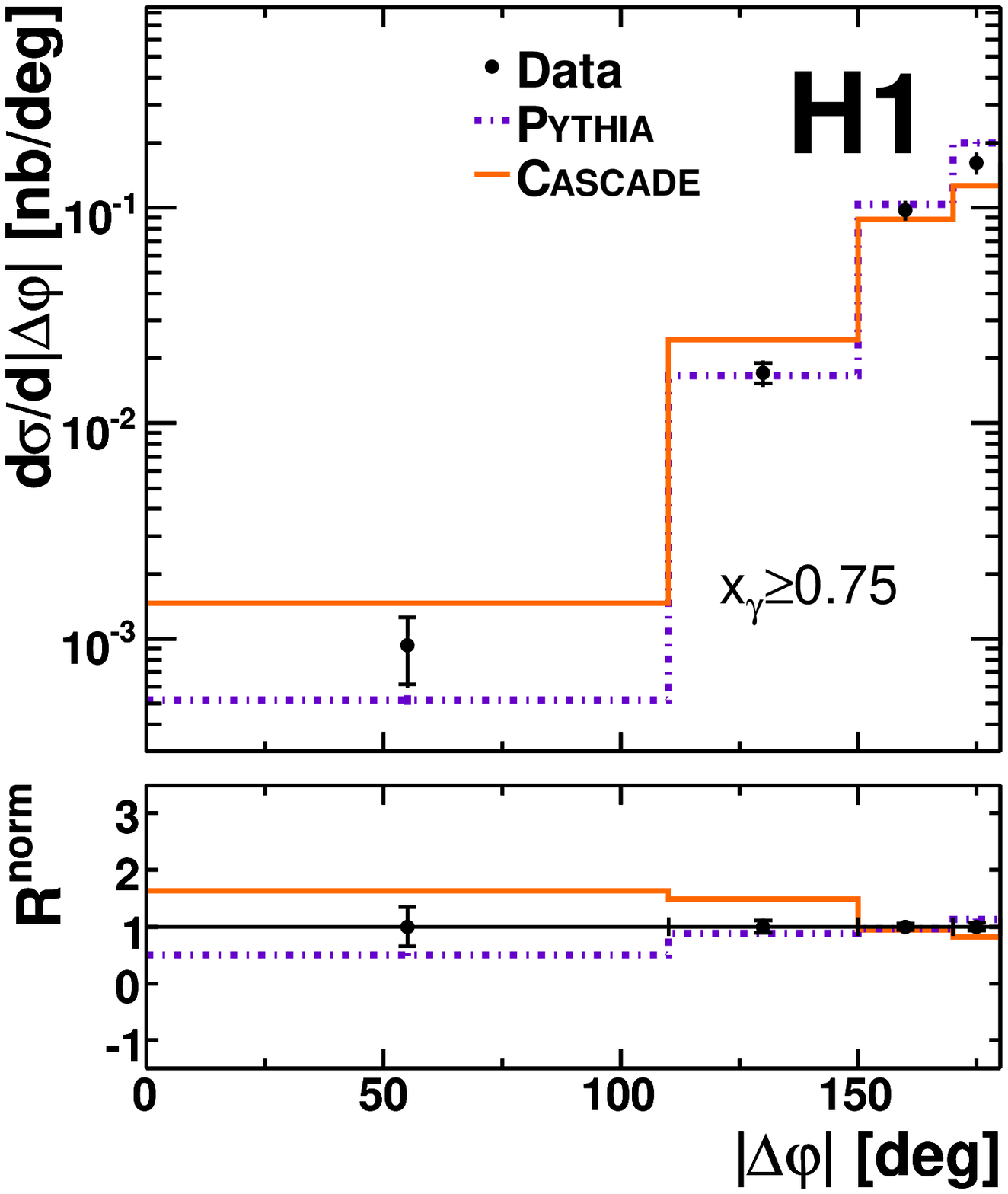}}
\put(0,-3){\includegraphics[width=8cm]{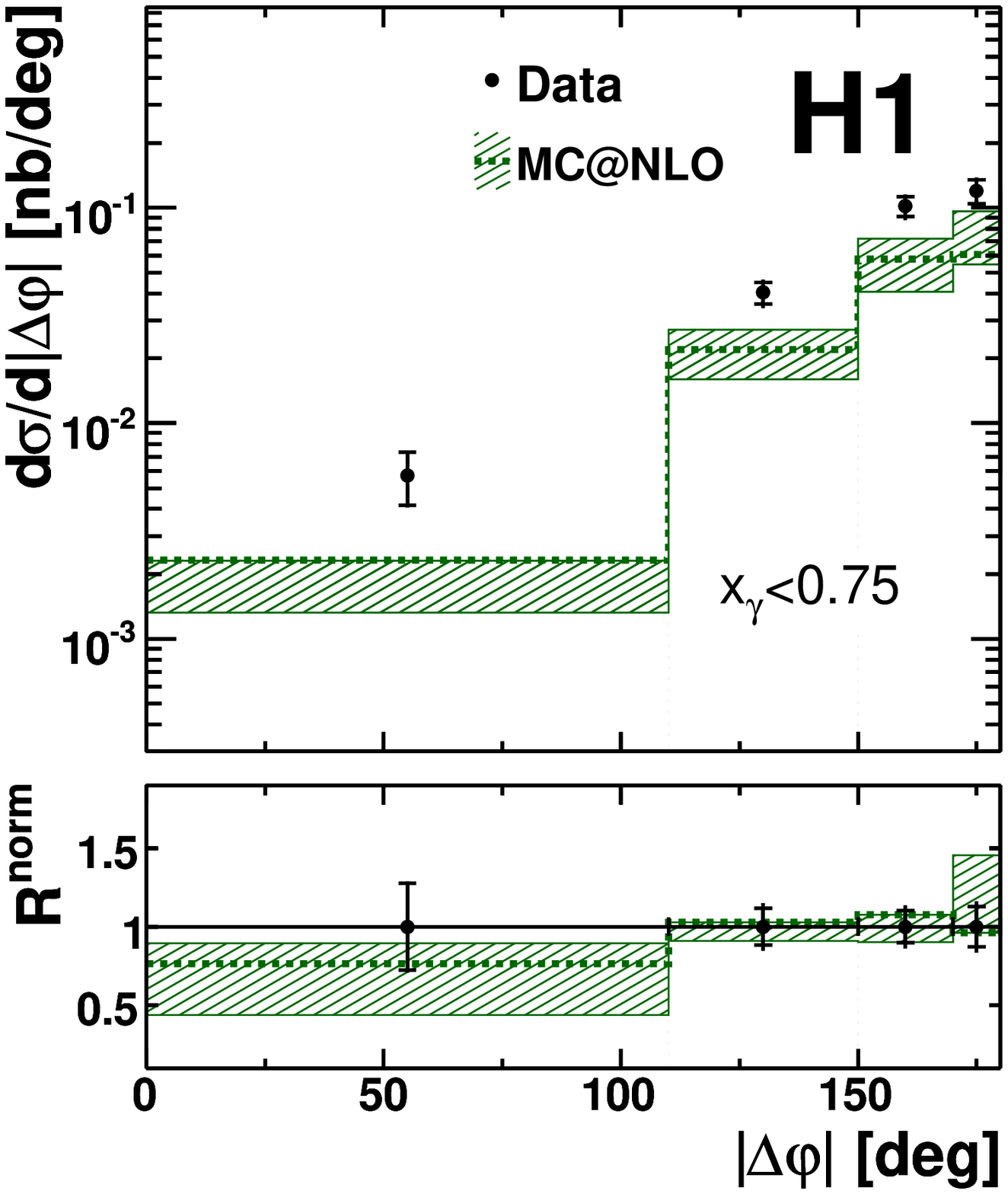}}
\put(8,-3){\includegraphics[width=8cm]{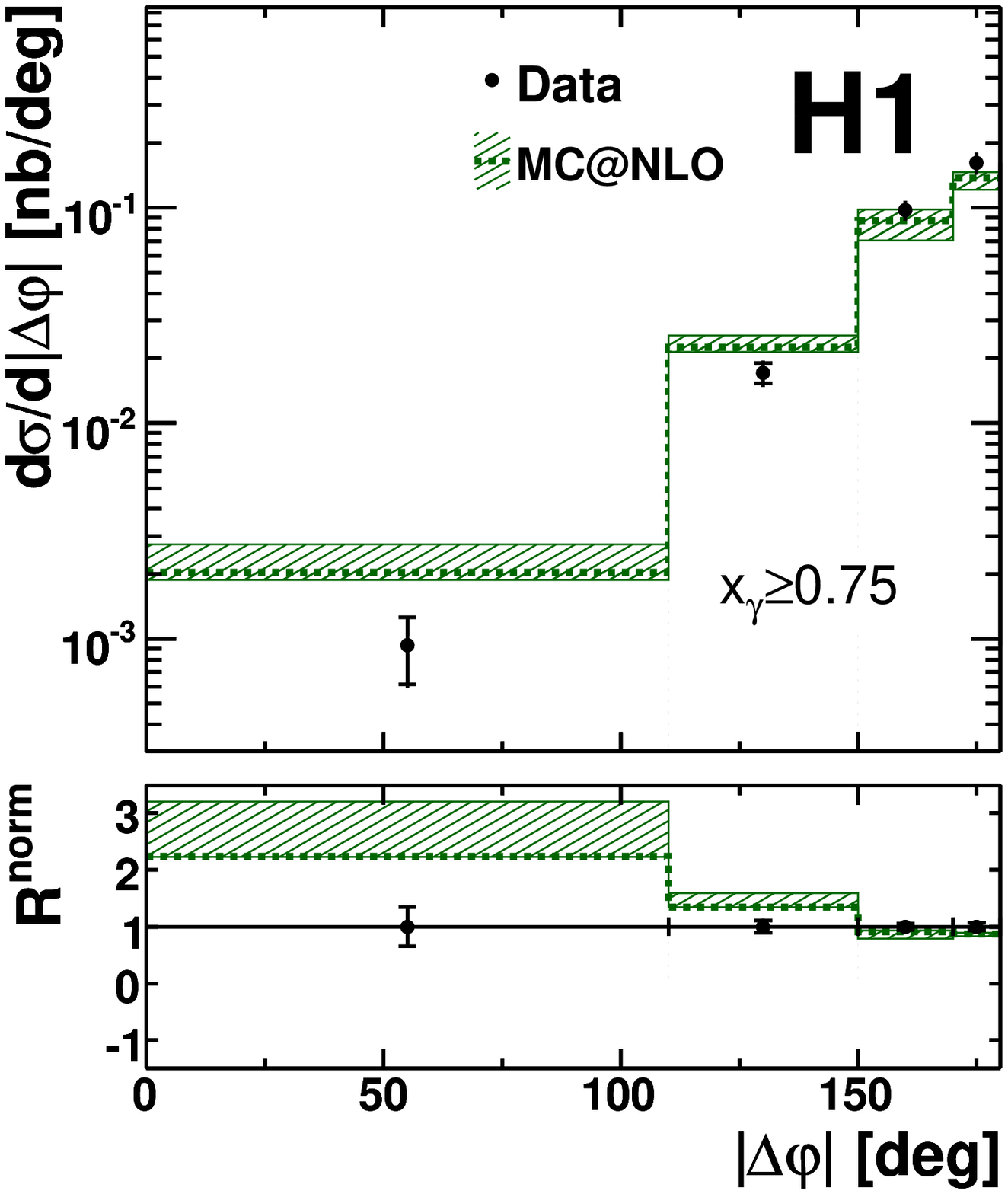}}
\put(1.8,17.5){\bf a)}
\put(9.8,17.5){\bf b)}
\put(1.8,8){\bf c)}
\put(9.8,8){\bf d)}
\end{picture}
   \caption{Single differential cross section for \dstarDj\ production as a function 
   of the difference in azimuthal angle $\Delta \varphi$ 
   in two regions of $\xgjj$  
    compared to predictions of \PYTHIA, \CASCADE\ and MC@NLO. 
    The normalised ratio \rnorm\ (see text) is also shown.}
    \label{fig:dstardijet-corr-2d}
\end{figure}

The \dphidsj\ dependence of the cross sections in two regions of $\xgjj$
is presented in table~\ref{tab:diffXsecXgammaDphi}
and in figure~\ref{fig:dstardijet-corr-2d}. \PYTHIA\ is in agreement
with the data.
\CASCADE\ overestimates the contribution from small \dphidsj\ in both 
$\xgjj$ regions. MC@NLO describes the shape well in the region of small
$\xgjj$, where resolved photon processes are enhanced, but is too low in
normalisation. At large $\xgjj$ values MC@NLO predicts the size of the cross section
correctly, but overestimates the contribution from small $|\Delta \varphi|$. 

The cross sections for \dstarDj\ production show that 
in general both hard partons in the final state can be described reasonably 
well by the QCD predictions, while the details
and especially the correlations between  the \dstarjet\ and the 
\otherj\ are not described very well by these theoretical calculations. 

\section{Conclusions}
The production of \dstar\ mesons in the photoproduction regime is 
investigated with the H1 detector 
at HERA with a seven times larger signal sample compared to the previous 
H1 measurement. The events containing \dstar\ mesons were triggered by
the tracks of the decay particles in the channel 
$D\*^{\pm}\rightarrow D^{0}\pi^{\pm}_{slow} \rightarrow 
K^{\mp}\pi^{\pm}\pi^{\pm}_{slow}$. Single and double differential cross 
sections are measured, and the results 
are compared to leading order QCD models provided by the MC simulation 
programs \PYTHIA\ and \CASCADE\ and to the next-to-leading order pQCD calculations 
FMNR, GMVFNS and MC@NLO.
The precision of the cross section measurements far exceeds the 
predictive power of the NLO theories. The shapes of the differential
cross sections, however, are less sensitive to the theoretical uncertainties,
and generally show reasonable agreement with the data.

The cross section for \dstarDj\ production is measured and compared to 
predictions of \PYTHIA, \CASCADE\ and MC@NLO.
The results are consistent with the expectation that the non-\dstar -jet
can originate not only from a charm quark but also from a light parton.
Significant contributions from higher order QCD radiation or transverse 
momenta of the partons in the initial state are needed
to describe the cross section away from the back-to-back configuration 
between the \dstarjet\ and \otherj\  at  
$\dphidsj \simeq 180 \grad$.
The cross sections as a function of the transverse momentum and 
the pseudorapidity 
of the \dstarjet\ and the \otherj\ are reasonably well described by 
the predictions. However, 
significant differences are observed in the description of 
some variables related to the structure of the hadronic final state, 
such as \dphidsj, $M_X$ and \xgjj . 

\section*{Acknowledgements}

We are grateful to the HERA machine group whose outstanding
efforts have made this experiment possible. 
We thank the engineers and technicians for their work in constructing and
maintaining the H1 detector, our funding agencies for 
financial support, the
DESY technical staff for continual assistance
and the DESY directorate for support and for the
hospitality which they extend to the non-DESY 
members of the collaboration.


\clearpage

\renewcommand{\arraystretch}{1.15} 
\begin{table}[tb!]
  \begin{center}
    \begin{tabular}{|rr|crr|}
      \hline
      \multicolumn{5}{|c|}{\bf \boldmath H1 inclusive \dstar\ cross sections} \\
      \hline
      \hline
      \multicolumn{2}{|c|}{\ptds\ range} & \dsdx{\ptds} & stat. & sys.  \\  
      \multicolumn{2}{|c|}{[GeV]} & \multicolumn{1}{c}{[nb/GeV]} &
      \multicolumn{1}{c}{[\%]} & \multicolumn{1}{c|}{[\%]} \\ 
      \hline
$ 1.8$ & $ 2.1 $ & $ 36   $ & $ \pm 12 $ & $\pm 13$  \\
$ 2.1$ & $ 2.5 $ & $ 29   $ & $ \pm 8  $ & $\pm 13$  \\
$ 2.5$ & $ 3.0 $ & $ 15   $ & $ \pm 5  $ & $\pm 11$  \\
$ 3.0$ & $ 3.5 $ & $ 8.6  $ & $ \pm 6  $ & $\pm  8$  \\
$ 3.5$ & $ 4.5 $ & $ 4.3  $ & $ \pm 3  $ & $\pm  8$  \\
$ 4.5$ & $ 5.5 $ & $ 2.3  $ & $ \pm 4  $ & $\pm  9$  \\
$ 5.5$ & $ 6.5 $ & $ 0.89 $ & $ \pm 5  $ & $\pm  7$  \\
$ 6.5$ & $ 9.0 $ & $ 0.25 $ & $ \pm 6  $ & $\pm  8$  \\
$ 9.0$ & $12.5 $ & $ 0.047$ & $ \pm 12 $ & $\pm 11$  \\
      \hline				             
      \hline
      \multicolumn{2}{|c|}{\etads\ range} & \dsdx{\etads} & stat. & sys. \\  
      \multicolumn{2}{|c|}{} & \multicolumn{1}{c}{[nb]} &
      \multicolumn{1}{c}{[\%]} & \multicolumn{1}{c|}{[\%]} \\ 
      \hline
$ -1.5$  & $-1.0 $ & $ 13 $ & $ \pm 5  $ & $ \pm 10$  \\
$ -1.0$  & $-0.5 $ & $ 16 $ & $ \pm 4  $ & $ \pm 10$  \\
$ -0.5$  & $ 0.0 $ & $ 18 $ & $ \pm 4  $ & $ \pm 10$  \\
$  0.0$  & $ 0.5 $ & $ 15 $ & $ \pm 4  $ & $ \pm 10$  \\
$  0.5$  & $ 1.0 $ & $ 12 $ & $ \pm 5  $ & $ \pm 10$  \\
$  1.0$  & $ 1.5 $ & $  7.9$& $ \pm 10 $ & $ \pm 10$  \\

      \hline
      \hline
      \multicolumn{2}{|c|}{\Wgp\ range} & \dsdx{(\Wgp)} & stat. & sys.\\  
      \multicolumn{2}{|c|}{[GeV]} & \multicolumn{1}{c}{[nb/G\eV]} &
      \multicolumn{1}{c}{[\%]} & \multicolumn{1}{c|}{[\%]} \\ 
      \hline
$ 100$  & $140 $ & $ 0.34 $ & $ \pm 3  $ & $ \pm 10$ \\
$ 140$  & $180 $ & $ 0.29 $ & $ \pm 3  $ & $ \pm 10$ \\
$ 180$  & $230 $ & $ 0.19 $ & $ \pm 4  $ & $ \pm 10$ \\
$ 230$  & $285 $ & $ 0.11 $ & $ \pm 6  $ & $ \pm 10$ \\
      \hline						      
      \hline
      \multicolumn{2}{|c|}{\zDs\ range} & \dsdx{(\zDs)} & stat. & sys.\\  
      \multicolumn{2}{|c|}{} & \multicolumn{1}{c}{[nb]} &
      \multicolumn{1}{c}{[\%]} & \multicolumn{1}{c|}{[\%]} \\ 
      \hline
$ 0.00$  & $0.10 $ & $ 45 $ & $ \pm 14 $ & $ \pm 11$ \\
$ 0.10$  & $0.20 $ & $ 89 $ & $ \pm 5  $ & $ \pm 11$  \\
$ 0.20$  & $0.35 $ & $ 76 $ & $ \pm 3  $ & $ \pm 10$  \\
$ 0.35$  & $0.55 $ & $ 55 $ & $ \pm 3  $ & $ \pm  9$  \\
$ 0.55$  & $1.00 $ & $ 13 $ & $ \pm 4  $ & $ \pm 11$  \\    
    \hline
    \end{tabular}
    \caption{Bin averaged single differential cross sections for inclusive \dstar\ 
    production in bins of \ptds, \etads, \Wgp\ and \zDs\
    with their statistical and uncorrelated systematic uncertainties. 
    The normalisation uncertainty of $6.0\%$ is not included.}
    \label{tab:cssd}
  \end{center}
\end{table}

\begin{table}[tb]
  \begin{center}
    \begin{tabular}{|rr|crr|}
      \hline
      \multicolumn{5}{|c|}{\bf \boldmath H1 inclusive \dstar\ cross sections} \\
      \hline
      \hline
      \multicolumn{5}{|c|}{$1.8 \leq \ptds < 2.5\ {\rm GeV}$} \\
      \hline	
      \multicolumn{2}{|c|}{\etads\ range} & 
      \multicolumn{1}{c}{${{\rm d}^2\sigma}/{{\rm d}  \eta}{{\rm d}  p_T}$} & 
      \multicolumn{1}{c}{stat.} & \multicolumn{1}{c|}{sys.}   \\ 
      \multicolumn{2}{|c|}{} & \multicolumn{1}{c}{[nb/GeV]} &
      \multicolumn{1}{c}{[\%]} & \multicolumn{1}{c|}{[\%]} \\        	
      \hline
$ -1.5 $ & $ -1.0 $ & $ 13 $ & $ \pm 12 $ & $ \pm 14$ \\
$ -1.0 $ & $ -0.5 $ & $ 12 $ & $ \pm 12 $ & $ \pm 14$ \\
$ -0.5 $ & $  0.0 $ & $ 14 $ & $ \pm 11 $ & $ \pm 13$ \\
$  0.0 $ & $  0.5 $ & $ 10 $ & $ \pm 16 $ & $ \pm 13$ \\
$  0.5 $ & $  1.5 $ & $  7.8$& $ \pm 18 $ & $ \pm 13$ \\
      \hline
      \hline
      \multicolumn{5}{|c|}{$2.5 \leq \ptds < 4.5\ {\rm GeV}$} \\
      \hline	
      \multicolumn{2}{|c|}{\etads\ range} & 
      \multicolumn{1}{c}{${{\rm d}^2\sigma}/{{\rm d}  \eta}{{\rm d}  p_T}$} & 
      \multicolumn{1}{c}{stat.} & \multicolumn{1}{c|}{sys.}   \\ 
      \multicolumn{2}{|c|}{} & \multicolumn{1}{c}{[nb/GeV]} &
      \multicolumn{1}{c}{[\%]} & \multicolumn{1}{c|}{[\%]} \\        	
      \hline
$ -1.5 $ & $ -1.0 $ & $ 2.2 $ & $ \pm 6 $ & $ \pm 9$  \\
$ -1.0 $ & $ -0.5 $ & $ 3.0 $ & $ \pm 4 $ & $ \pm 9$  \\
$ -0.5 $ & $  0.0 $ & $ 3.6 $ & $ \pm 5 $ & $ \pm 9$  \\
$  0.0 $ & $  0.5 $ & $ 3.0 $ & $ \pm 5 $ & $ \pm 9$  \\
$  0.5 $ & $  1.0 $ & $ 2.3 $ & $ \pm 7 $ & $ \pm 9$  \\
$  1.0 $ & $  1.5 $ & $ 1.8 $ & $\pm 14 $ & $ \pm 9$  \\
\hline
      \hline
      \multicolumn{5}{|c|}{$4.5 \leq \ptds < 12.5\ {\rm GeV}$} \\
      \hline	
      \multicolumn{2}{|c|}{\etads\ range} & 
      \multicolumn{1}{c}{${{\rm d}^2\sigma}/{{\rm d}  \eta}{{\rm d}  p_T}$} & 
      \multicolumn{1}{c}{stat.} & \multicolumn{1}{c|}{sys.}   \\ 
      \multicolumn{2}{|c|}{} & \multicolumn{1}{c}{[nb/GeV]} &
      \multicolumn{1}{c}{[\%]} & \multicolumn{1}{c|}{[\%]} \\        	
      \hline
$ -1.5 $ & $ -1.0 $ & $ 0.070$ & $ \pm 10 $ & $ \pm 12$  \\
$ -1.0 $ & $ -0.5 $ & $ 0.14 $ & $ \pm 6  $ & $ \pm 11$  \\
$ -0.5 $ & $  0.0 $ & $ 0.22 $ & $ \pm 6  $ & $ \pm 11$  \\
$  0.0 $ & $  0.5 $ & $ 0.24 $ & $ \pm 5  $ & $ \pm 11$  \\
$  0.5 $ & $  1.0 $ & $ 0.18 $ & $ \pm 6  $ & $ \pm 11$  \\
$  1.0 $ & $  1.5 $ & $ 0.11 $ & $ \pm 10 $ & $ \pm 12$  \\ 
      \hline
    \end{tabular}
    \caption{Bin averaged double differential cross sections for inclusive \dstar\ 
    production in bins of \etads\
    for three ranges in \ptds\ with their statistical and uncorrelated
    systematic uncertainties.  
    The normalisation uncertainty of $6.0\%$ is not included.}
    \label{tab:csdd}
  \end{center}
\end{table}

\begin{table}[tb!]
  \begin{center}
    \begin{tabular}{|crrc|crr|}
      \hline
      \multicolumn{7}{|c|}{\bf \boldmath H1 \dstarDj\ cross sections} \\
      \hline
      \hline
      \multicolumn{4}{|c|}{$\eta(\dstarjet)$ range} & \dsdx{\eta(\dstarjet)} & stat. & sys.  \\  
      \multicolumn{4}{|c|}{} & \multicolumn{1}{c}{[nb]} &
      \multicolumn{1}{c}{[\%]} & \multicolumn{1}{c|}{[\%]} \\ 
      \hline
 & $ -1.5 $ & $ -1.0 $ &  & $ 2.3 $ & $ \pm 12 $ & $ \pm 11 $ \\
 & $ -1.0 $ & $ -0.5 $ &  & $ 3.2 $ & $ \pm 7  $ & $ \pm  8 $ \\
 & $ -0.5 $ & $  0.0 $ &  & $ 3.9 $ & $ \pm 7  $ & $ \pm  8 $ \\
 & $  0.0 $ & $  0.5 $ &  & $ 3.9 $ & $ \pm 8  $ & $ \pm  8 $ \\
 & $  0.5 $ & $  1.0 $ &  & $ 3.4 $ & $ \pm 9  $ & $ \pm  8 $ \\
 & $  1.0 $ & $  1.5 $ &  & $ 2.8 $ & $ \pm 14 $ & $ \pm  8 $ \\
\hline
      \hline
      \multicolumn{4}{|c|}{$\eta(\otherj)$ range} & \dsdx{\eta(\otherj)} & stat. & sys.  \\  
      \multicolumn{4}{|c|}{} & \multicolumn{1}{c}{[nb]} &
      \multicolumn{1}{c}{[\%]} & \multicolumn{1}{c|}{[\%]} \\ 
      \hline
 & $ -1.5 $ & $ -1.0 $ &  & $ 1.2 $ & $ \pm 15 $ & $ \pm 11 $ \\
 & $ -1.0 $ & $ -0.5 $ &  & $ 1.3 $ & $ \pm 13 $ & $ \pm  9 $ \\
 & $ -0.5 $ & $  0.0 $ &  & $ 2.1 $ & $ \pm 10 $ & $ \pm  8 $ \\
 & $  0.0 $ & $  0.5 $ &  & $ 2.6 $ & $ \pm 9  $ & $ \pm  8 $ \\
 & $  0.5 $ & $  1.0 $ &  & $ 2.7 $ & $ \pm 8  $ & $ \pm  8 $ \\
 & $  1.0 $ & $  1.5 $ &  & $ 2.9 $ & $ \pm 8  $ & $ \pm  8 $ \\
 & $  1.5 $ & $  2.2 $ &  & $ 2.5 $ & $ \pm 10 $ & $ \pm  8 $ \\
 & $  2.2 $ & $  2.9 $ &  & $ 2.2 $ & $ \pm 15 $ & $ \pm  8 $ \\
\hline
      \hline
      \multicolumn{4}{|c|}{\pt(\dstarjet)\ range} & \dsdx{\pt(\dstarjet)} & stat. & sys.  \\  
      \multicolumn{4}{|c|}{[GeV]} & \multicolumn{1}{c}{[nb/GeV]} &
      \multicolumn{1}{c}{[\%]} & \multicolumn{1}{c|}{[\%]} \\ 
      \hline
 & $ 3.5 $ & $ 5.0 $ &  & $ 2.7 $ & $ \pm 8 $ &  $ \pm  8 $ \\
 & $ 5.0 $ & $ 8.0 $ &  & $ 1.4 $ & $ \pm 5 $ &  $ \pm  7 $ \\
 & $ 8.0 $ & $15.0 $ &  & $ 0.17$ & $ \pm 7 $ &  $ \pm  7 $ \\
\hline
      \hline
      \multicolumn{4}{|c|}{\pt(\otherj)\ range} & \dsdx{\pt(\otherj)} & stat. & sys.  \\  
      \multicolumn{4}{|c|}{[GeV]} & \multicolumn{1}{c}{[nb/GeV]} &
      \multicolumn{1}{c}{[\%]} & \multicolumn{1}{c|}{[\%]} \\ 
      \hline
 & $ 3.5 $ & $ 5.0 $ &  & $ 3.0 $ & $ \pm 7 $ &  $ \pm 8 $ \\
 & $ 5.0 $ & $ 8.0 $ &  & $ 1.2 $ & $ \pm 5 $ &  $ \pm 7 $ \\
 & $ 8.0 $ & $15.0 $ &  & $ 0.24 $ & $ \pm 7 $ & $ \pm 10 $ \\
      \hline
    \end{tabular}
    \caption{Bin averaged single differential cross sections for \dstarDj\ production
    in bins of $\eta$ and \pt\ of the \dstarjet\ and the \otherj\
    with their statistical and uncorrelated systematic uncertainties.  
    The normalisation uncertainty of $6.0\%$ is not included.}
    \label{tab:diffXsecdijet}
  \end{center}
\end{table}

\begin{table}[tb!]
  \begin{center}
    \begin{tabular}{|rr|crr|}
      \hline
      \multicolumn{5}{|c|}{\bf \boldmath H1 \dstarDj\ cross sections} \\
      \hline
      \hline
      \multicolumn{2}{|c|}{$\Delta \eta$ range} & \dsdx{\Delta \eta} & stat. & sys.  \\  
      \multicolumn{2}{|c|}{} & \multicolumn{1}{c}{[nb]} &
      \multicolumn{1}{c}{[\%]} & \multicolumn{1}{c|}{[\%]} \\ 
      \hline
$ -3.0 $ & $ -2.0 $ & $ 0.24 $ & $ \pm 33 $ & $ \pm 13 $ \\ 
$ -2.0 $ & $ -1.0 $ & $ 0.85 $ & $ \pm 12 $ & $ \pm  9 $ \\ 
$ -1.0 $ & $  0.0 $ & $ 1.7  $ & $ \pm  9 $ & $ \pm  8 $ \\ 
$  0.0 $ & $  1.0 $ & $ 2.4  $ & $ \pm  7 $ & $ \pm  8 $ \\ 
$  1.0 $ & $  2.0 $ & $ 2.5  $ & $ \pm  7 $ & $ \pm  8 $ \\ 
$  2.0 $ & $  3.0 $ & $ 1.6  $ & $ \pm 11 $ & $ \pm  8 $ \\ 
$  3.0 $ & $  4.0 $ & $ 0.63 $ & $ \pm 21 $ & $ \pm 12 $ \\ 
$  4.0 $ & $  4.4 $ & $ 0.22 $ & $ \pm 79 $ & $ \pm 31 $ \\ 
\hline
      \hline
      \multicolumn{2}{|c|}{$|\Delta \varphi|$ range} & \dsdx{|\Delta \varphi|} & stat. & sys.  \\  
      \multicolumn{2}{|c|}{[deg.]} & \multicolumn{1}{c}{[nb/deg.]} &
      \multicolumn{1}{c}{[\%]} & \multicolumn{1}{c|}{[\%]} \\ 
      \hline
$   0 $ & $ 110 $ & $ 0.0066 $ & $ \pm 24$ &$ \pm 8 $ \\ 
$ 110 $ & $ 150 $ & $ 0.057 $ & $ \pm 8 $ & $ \pm 8 $ \\ 
$ 150 $ & $ 170 $ & $ 0.20 $ & $ \pm 5 $  & $ \pm 7 $ \\ 
$ 170 $ & $ 180 $ & $ 0.28 $ & $ \pm 6 $  & $ \pm 8 $ \\ 
\hline
      \hline
      \multicolumn{2}{|c|}{$M_X$ range} & \dsdx{M_X} & stat. & sys.  \\  
      \multicolumn{2}{|c|}{[GeV]} & \multicolumn{1}{c}{[nb/GeV]} &
      \multicolumn{1}{c}{[\%]} & \multicolumn{1}{c|}{[\%]} \\ 
      \hline
$  30 $ & $  75 $ & $ 0.075 $ & $ \pm 4 $ & $ \pm 7 $ \\ 
$  75 $ & $ 120 $ & $ 0.069 $ & $ \pm 7 $ & $ \pm 7 $ \\ 
$ 120 $ & $ 250 $ & $ 0.024 $ & $ \pm 11 $& $ \pm 7 $ \\ 
\hline
     \hline
      \multicolumn{2}{|c|}{\xgjj\ range} & \dsdx{\xgjj} & stat. & sys.  \\  
      \multicolumn{2}{|c|}{} & \multicolumn{1}{c}{[nb]} &
      \multicolumn{1}{c}{[\%]} & \multicolumn{1}{c|}{[\%]} \\ 
      \hline
$ 0.00 $ & $ 0.45 $ & $ 4.9 $ & $ \pm 15 $ & $ \pm 9 $ \\ 
$ 0.45 $ & $ 0.75 $ & $ 11 $ & $ \pm 7 $ &   $ \pm 8 $ \\ 
$ 0.75 $ & $ 1.00 $ & $ 17 $ & $ \pm 4 $ &   $ \pm 7 $ \\ 
      \hline
     \end{tabular}
    \caption{Bin averaged single differential cross sections for \dstarDj\ production
    in bins of $\Delta \eta$, $|\Delta \varphi|$, \xgjj\ and $M_X$ 
    with their statistical and uncorrelated systematic uncertainties.  
    The normalisation uncertainty of $6.0\%$ is not included.}
    \label{tab:diffXsecdijetcorr}
  \end{center}
\end{table}

\begin{table}[tb!]
  \begin{center}
    \begin{tabular}{|rr|crr|}
      \hline
      \multicolumn{5}{|c|}{\bf \boldmath H1 \dstarDj\ cross sections} \\
      \hline
      \hline
      \multicolumn{5}{|c|}{$\xgjj < 0.75$} \\
      \multicolumn{2}{|c|}{$|\Delta \varphi|$ range} & \dsdx{|\Delta \varphi|} & stat. & sys.  \\  
      \multicolumn{2}{|c|}{[deg.]} & \multicolumn{1}{c}{[nb/deg.]} &
      \multicolumn{1}{c}{[\%]} & \multicolumn{1}{c|}{[\%]} \\ 
      \hline
  $ 0 $ & $ 110 $ & $ 0.0057 $ & $ \pm 28 $ & $ \pm 9 $ \\ 
$ 110 $ & $ 150 $ & $ 0.040 $ & $ \pm 12 $ & $ \pm 9 $ \\ 
$ 150 $ & $ 170 $ & $ 0.10 $ & $ \pm 10 $ & $ \pm 9 $ \\ 
$ 170 $ & $ 180 $ & $ 0.12 $ & $ \pm 13 $ & $ \pm 10 $ \\ 
\hline
      \hline
      \multicolumn{5}{|c|}{$\xgjj \ge 0.75$} \\
      \multicolumn{2}{|c|}{$|\Delta \varphi|$ range} & \dsdx{|\Delta \varphi|} & stat. & sys.  \\  
      \multicolumn{2}{|c|}{[deg.]} & \multicolumn{1}{c}{[nb/deg.]} &
      \multicolumn{1}{c}{[\%]} & \multicolumn{1}{c|}{[\%]} \\ 
      \hline
  $ 0 $ & $ 110 $ & $ 0.0009 $ &$ \pm 34 $ & $ \pm 12 $ \\
$ 110 $ & $ 150 $ & $ 0.017 $ & $ \pm 11 $ & $ \pm  8 $ \\
$ 150 $ & $ 170 $ & $ 0.097 $ & $ \pm 6  $ & $ \pm  8 $ \\
$ 170 $ & $ 180 $ & $ 0.16 $ &  $ \pm 6  $ & $ \pm  9 $ \\
\hline
     \end{tabular}
    \caption{Bin averaged single differential cross sections for \dstarDj\ production
    in bins of $|\Delta \varphi|$ in two regions of \xgjj\ 
    with their statistical and uncorrelated systematic uncertainties.  
    The normalisation uncertainty of $6.0\%$ is not included.}
    \label{tab:diffXsecXgammaDphi}
  \end{center}
\end{table}

\newpage

\newpage

\newpage

\newpage
\newpage

\newpage

\newpage

\newpage

\newpage

\newpage

\end{document}